\documentclass[a4paper]{article}

\usepackage{graphicx}
\usepackage[longnamesfirst]{natbib}
\shortcites{Rosenblum2011}
\shortcites{mirouh2012}
\shortcites{Castaing1989}
\shortcites{Timmermanns2008}

\usepackage{amssymb}
\usepackage{amsbsy}
\usepackage{authblk}

\providecommand\bnabla{\boldsymbol{\nabla}}
\providecommand\bcdot{\boldsymbol{\cdot}}

\newcommand\Pran{\ensuremath{Pr}} 
\newcommand\Ta{\ensuremath{Ta}}   
\newcommand\Ra{\ensuremath{Ra}}   
\newcommand\Tac{\ensuremath{\Ta_{\mathrm{crit}}}}   
\newcommand\Le{\ensuremath{Le}}   
\newcommand\Nu{\ensuremath{Nu}}   
\newcommand\Ro{\ensuremath{Ro}}   
\newcommand\Roeff{\ensuremath{\Ro_{\mathrm{eff}}}}   
\newcommand\Rat{\ensuremath{\Ra_T}}   
\newcommand\Nut{\ensuremath{\Nu_T}}   
\newcommand\Nus{\ensuremath{\Nu_S}}   

\begin{document}

\title{The effects of rotation on a double-diffusive layer in a rotating spherical shell}

\author[1]{P. M. Blies \thanks{Email address for correspondence: patrick.blies@univie.ac.at}}
\author[1]{F. Kupka}
\author[2]{F. Zaussinger}
\author[3]{R. Hollerbach}

\affil[1]{Department of Mathematics, University of Vienna, Oskar-Morgenstern-Platz 1, 1090 Vienna, Austria}
\affil[2]{Department of Aerodynamics and Fluid Mechanics, Brandenburg University of Technology, Siemens-Halske-Ring 14, 03046 Cottbus, Germany}
\affil[3]{Department of Applied Mathematics, University of Leeds, Leeds LS2 9JT, U.K.}
\maketitle

\begin{abstract}
	So far, numerical studies of double-diffusive layering in turbulent convective flows 
	have neglected the effects of rotation. We undertake a first step into that direction by investigating how Coriolis forces affect a double-diffusive layer inside a
	rotating spherical shell. For this purpose we have run simulations in a parameter regime where these layers are expected to form and successively increased the rate of rotation with the result that fast rotation is found
	to have a similar stabilising effect on the overall convective flux as an increase of the stability ratio $R_\rho$ has in a non-rotating setup. We have also studied to what extent the regimes of rotational constraints suggested
	by \citet{king2013} for rotation in the case of Rayleigh-B\'enard convection are influenced by double-diffusive convection: their classification could also be applicable to the case of double-diffusive convection in a
	spherical shell if it is extended to be also a function of the stability ratio $R_\rho$. Furthermore, we examined the ratio of saline and thermal Nusselt numbers and compared our results with models of
	\citet{spruit_theory_2013}, \citet{Rosenblum2011} and \citet{wood_2013}.  We find our data to be fitted best by Spruit's model. Our result that fast rotation further decreases the convective transport, which is already
	lowered by double-diffusive convection, could play a major role for e.g. the modeling of the interior of some rapidly rotating giant planets, as gaseous giant planets have recently been proposed to be influenced by
	double-diffusive convection.
\end{abstract}

\section{Introduction}

The physical process known as double-diffusive convection was first described in the 1950's by \citet{stommel1956} who observed the effect in an experiment. Shortly afterwards, it was also found in astrophysics when 
the first detailed stellar models were computed and \citet{schwarzschild_haerm_1958} found irregularities in their calculations concerning whether or not a zone with a gradient in molecular weight was 
stable according to the Ledoux criterion or the Schwarzschild criterion.
But even more than fifty years after its discovery, the field is still actively researched which is due to two reasons: on the one hand it lacked immediate practical incentives that have accelerated the 
development of other branches of fluid mechanics. On the other hand, numerical simulations were not possible for a long time because of the considerable computational expenses they demand.
For a summary of the historical development of the area see the paper by \citet{huppert_1981}, for a recent physical review about semiconvection see \citet{zaussinger_kupka_muthsam_2012}. \\

Double-diffusive convection occurs in situations where the effect of a thermal gradient on stability and the effect of a molecular weight gradient on stability compete with each other: if the temperature gradient stabilises the
system and the molecular weight gradient destabilises it, thermohaline convection can occur. Its distinguishing property is the appearance of flow structures known as salt-fingers (thus also the name salt-fingering convection).
In the opposite case (temperature gradient unstable and molecular weight gradient stable) layering convection/semiconvection can occur. 
Note that we used the term ``can occur''. Whether thermohaline/layering convection really does occur
depends on the ratio of the molecular weight buoyancy frequency to the thermal buoyancy frequency, the so called stability ratio $R_\rho=-N^2_{\mu}/N^2_{T}$. In the incompressible case it is equivalent to the ratio of the Rayleigh
numbers associated with the thermal instability and the instability caused by the molecular weight.  In this paper, our focus will be on layering convection. Situations where this process occurs on earth include the convection in
the arctic ocean where cool and fresh melt water from above leads to a destabilising negative temperature gradient and a stabilising negative molecular gradient in salt \citep{turner2010}. Other examples are East-African rift
lakes which are heated from below by volcanic activity. This leads to a temperature gradient (unstable) and
causes dissolved gases like methane and carbon dioxide to be introduced into the system, thus causing a stabilising molecular weight gradient \citep{Schmid2010225}.

But not only systems on earth are prone to double-diffusive convection: it can also occur in astrophysical systems like in icy satellites, giant planets and massive stars.
Very recently, \citet{ORourke2014} have investigated the effects of a stabilising compositional gradient and the resulting double-diffusive convection in Titan.
The role of semiconvection for the interior of giant planets 
has been discussed by \citet{stevenson1982a} and recently by \citet{chabrier2007} who suggested that it might be responsible for 
the radius anomalies of some hot Jupiters. This thought is further developed by \citet{Leconte2012}; they investigated the effect of semiconvection on the interior structure of planets and showed that it 
could explain the luminosity anomaly of Saturn \citep{Leconte2013}. 
They also point out: ``Determining the solute transport properties in the regime of
layered convection more precisely, however, will be central to evolutionary calculations. 3D hydrodynamical simulations in a realistic parameter range are thus strongly needed.'' \citep{Leconte2012}. 

However, numerical simulations of double-diffusive systems in a realistic astrophysical parameter range pose a serious challenge (and are, in fact, still impossible in the stellar regime with today's computers) 
because of the huge spread of length and time scales of which the smallest length scale --- the size of the diffusive boundary layer --- needs to be resolved. For example, the ratio of the diffusivities of temperature
and solute (the Lewis number) for the plasma
in the interior of a semiconvective region of a star is $ \Le \approx 10^{-9}$ \citep{zaussinger_diss}. This would require an impossible spatial resolution if one were to attempt a DNS of such a zone \citep[also][]{zaussinger_diss}.
While simulations are nowhere near the realistic parameter range for stellar astrophysical conditions yet, the parameter regime of giant planets has become feasible with today's computers since their Prandtl and Lewis numbers are
much more moderate: the Prandtl number ranges from $\Pran = 10^{-2}$ to $1$, the Lewis number is about $\Le = 0.01$ \citep{chabrier2007}.
For idealised microphysics, there are a number of simulations in two dimensions \citep[e.g.][]{zaussinger_scn_2013} and in three dimensions \citep[e.g.][]{wood_2013} in this parameter regime.
Recently, a simulation in a realistic parameter range for the 
Atlantic Ocean that correctly reproduces measurements has been conducted by \citet{Flanagan20132466}. 

However, all of the mentioned studies have neglected the effect of rotation on the development of double-diffusive convection.
While this may be justifiable in the case of thin layers as they are occuring in the Arctic ocean (layer thickness 1 to 5 m, see \citealt{Timmermanns2008}) or in lake Kivu (average thickness of the mixed layers 0.48 m, 
see \citealt{Schmid2010225}),
it might not be negligible for large layers that could be forming in global convection zones on rapidly rotating giant planets and stars. 

It might even prove to be essential if trying to determine if layered convection is indeed occurring in giant planets and stars and what its precise
influences on the transport properties are. Our work is a first step
in the direction of investigating the effects of rotation on semiconvective layers. We note that while \citet{Net2012} did study thermosolutial 
convection in rotating spherical shells, they investigated a different parameter regime than the one where layers are expected to form
so their work gives us no lead as to how semiconvective \textit{layers} are influenced by rotation. 

We want to give a remark on nomenclature here:
in oceans the molecular weight gradient is caused by dissolved salt. That is why salinity
gradient is another common term for the molecular weight gradient, particularly in oceanography. We will use the term salinity in this papers as well because it is handier than molecular weight of second species. We
assume salinity to be the concentration of the solute, no matter what exactly the solute is.

The publication is structured as follows: in chapter \ref{sec:model_description} we present the physical model and the underlying equations. We introduce the governing dimensionless numbers and the boundary conditions.
In chapter \ref{sec:numerical_implementation} we discuss the numerical setup. In chapter \ref{sec:results} we present the results of our simulations. First, we show the results for a run with 
one set of parameters without rotation to have a reference framework to which we can compare the following runs (chapter \ref{sec:nonrotating}). Next, we present the results of the simulations with rotation
(chapter \ref{sec:rotation}) and highlight some differences before investigating the influence of a change of the Prandtl number $\Pran$ and the density ratio $R_\rho$ chapter \ref{sec:modifying_pr_and_rrho}. 
This is followed by a discussion in chapter \ref{sec:discussion} and conclusions in chapter \ref{sec:conclusion}.

\section{Model description}\label{sec:model_description}
\subsection{The governing equations}
Two concentric spherical shells are maintained at different, constant temperatures: 
a hot inner sphere (radius $R_1$, temperature $T_1$) and a cool outer sphere (radius $R_2$, temperature $T_2$). The axis of rotation is taken to be the $z$-axis, 
the rate of rotation is constant and parallel to the z-direction: $\boldsymbol{\Omega} = \Omega \boldsymbol{e_z}$. Gravity operates inwards in radial direction: $\boldsymbol{g}= - g \boldsymbol{e_r}$.
The main simulation was run over a simulation time of almost one full 
thermal diffusion time scale. We studied the effects of an increase of the rate of rotation on the temporal evolution of a double-diffusive state.

For constant viscosities and diffusivities, the Navier--Stokes equations in the Boussinesq form including rotation and conservation of solute, read
\begin{equation}
	\bnabla \bcdot \boldsymbol{u} = 0,
	\label{eq:ns1}
\end{equation}
\begin{equation}
	\left( \frac{\partial \boldsymbol{u} }{\partial t} \right) + (\boldsymbol{u} \bcdot \bnabla) \boldsymbol{u} = - \frac{\bnabla p}{\rho_0} + \nu \nabla^{2} \boldsymbol{u} + \frac{\rho}{\rho_0} \boldsymbol{g} 
	- 2 \boldsymbol{\Omega} \times \boldsymbol{u} - \boldsymbol{\Omega} \times (\boldsymbol{\Omega} \times \boldsymbol{r}),
	\label{eq:ns2}
\end{equation}
\begin{equation}
	\frac{\partial T}{\partial t} + (\boldsymbol{u} \bcdot \bnabla) T = \kappa_T \nabla^{2} T,
	\label{eq:ns3}
\end{equation}
\begin{equation}
	\frac{\partial S}{\partial t} + (\boldsymbol{u} \bcdot \bnabla) S = \kappa_S \nabla^{2} S,
	\label{eq:ns4}
\end{equation}

with $\rho = \rho_0 [1 - \alpha (T-T_0) + \beta ( S- S_0)] $.\\

$\boldsymbol{u}$ is the velocity of the flow, $p$ the pressure, $\rho$ the density, 
$\rho_0$ a reference density, $\nu$ the kinematic viscosity, $\boldsymbol{\Omega}$ the angular velocity, $\boldsymbol{r}$ the position vector, $T$ the temperature, $T_0$ the reference temperature where $\rho = \rho_0$,
$S$ the salinity, $S_0$ the reference salinity where $\rho = \rho_0$, $\kappa_T$ the thermal diffusivity and $\kappa_S$ 
the molecular diffusivity. $\alpha$ is the thermal expansion 
coefficient $-\rho_0^{-1}  \, (\partial \rho/\partial T)_S$, $\beta$ is the saline expansion coefficient $\rho_0^{-1} \, (\partial \rho / \partial S)_T$.
With a later application in astrophysics in mind, we concentrated on systems where the centrifugal force is assumed much smaller than the gravitational force.
Hence, the term describing centrifugal forces $(- \boldsymbol{\Omega} \times (\boldsymbol{\Omega} \times \boldsymbol{r}))$
 will be neglected.

The equations are nondimensionalized with the scales
\begin{eqnarray}
	r &=& L r^*, \quad T - T_0 = \Delta T T^*, \quad S - S_0 = \Delta S S^*, \quad \boldsymbol{u} = \frac{\kappa_T}{L} \boldsymbol{u^*}, \quad t = \frac{L^2}{\kappa_T} t^*.
	\label{eq:nondim}
\end{eqnarray}

$L=R_2 - R_1$ is the difference between outer and inner radius, $\Delta T$ and $\Delta S$ are the differences of temperature and salinity between outer and inner radius. The time scale used is the thermal diffusion time scale.
Inserting (\ref{eq:nondim}) into (\ref{eq:ns1}) -- (\ref{eq:ns4}) and dropping the asterisks leads to the dimensionless form of the equations:
\begin{equation}
	\bnabla \bcdot \boldsymbol{u} = 0,
\end{equation}
\begin{equation}
	\Pran^{-1} \left[ \left( \frac{\partial \boldsymbol{u} }{\partial t} \right) + (\boldsymbol{u} \bcdot \bnabla) \boldsymbol{u} \right] = - \bnabla p_{\mathrm{eff}} + \nabla^{2} \boldsymbol{u} +
	Ra_T T \boldsymbol{e_r} - Ra_S S \boldsymbol{e_r} - \sqrt{\Ta} \boldsymbol{e_z} \times \boldsymbol{u},
\end{equation}
\begin{equation}
	\frac{\partial T}{\partial t} + (\boldsymbol{u} \bcdot \bnabla) T = \nabla^2 T,
\end{equation}
\begin{equation}
	\frac{\partial S}{\partial t} + (\boldsymbol{u} \bcdot \bnabla) S = \Le \, \nabla^2 S,
\end{equation}
where we introduced the dimensionless numbers 
\begin{eqnarray*}
	\Pran &=& \frac{\nu}{\kappa_T}, \quad	\Le = \frac{\kappa_S}{\kappa_T}, \quad Ra_T = \frac{\alpha L^3 \Delta T g }{ \kappa_T \nu}, \quad Ra_S = \frac{\beta L^3 \Delta S g }{ \kappa_T \nu},
	\quad \Ta=\frac{4 \Omega^2 L^4 }{ \nu^2}.
\end{eqnarray*}
$\Pran$ is the Prandtl number, $\Le$ is the Lewis number, $Ra_{T/S}$ are thermal and saline Rayleigh numbers, respectively, and $\Ta$ is the Taylor number. 
$Ra_T$ and $Ra_S$ are related to each other by the stability parameter $R_\rho=Ra_S/Ra_T$. Note that $t$ now stands for the time in thermal diffusion time scales, so that $t=1$ means that one thermal
diffusion time scale has passed. We also introduced the effective pressure $p_\mathrm{eff}$ which is the gradient of the scaled original pressure 
plus a constant term. This can be written in this form because the constant term will vanish in the course of solving the equations.

\subsection{Boundary and initial conditions, parameter regime} 

The idea behind our setup is to observe a growing double-diffusive layer and to investigate how it is influenced by rotation. To achieve that, we chose the following boundary and initial conditions.
\subsubsection{Boundary conditions}
We applied no-slip boundary conditions which read 
\[ \boldsymbol{u}(R_1) = 0, \quad \boldsymbol{u}(R_2) = 0, \quad T(R_1) = 1, \quad T(R_2) = 0, \quad S(R_1) = 1, \quad S(R_2) = 0. \]
These are reasonable boundary conditions because we assume our spherical shell to be one layer of a so called double-diffusive stack. The appropriateness of these boundary conditions is explained in 
chapter 3.2 of \citet{zaussinger_scn_2013}.

\subsubsection{Initial conditions}
The same assumption (taking the shell to be one layer of many) demands a step-like initial distribution of temperature and salt. However, since the purpose of this work is to investigate the effects of 
rotation on a developed double-diffusive layer it was imperative for a layer to form within our simulated spherical shell. We ran some simulations with different initial conditions: step-like initial distributions 
of both temperature and salinity and a step-like distribution of one quantity and a linear distribution of the other one. Our goal was to test the influence of these different initial conditions on the thermal Nusselt number.
The thermal and saline Nusselt numbers are measures for the convective heat and salt flux at the boundaries of the system, respectively. For incompressible flows (which we are looking at) 
they are defined as the ratio of the total heat or salt flux and the heat or salt flux that would be transported by conduction alone:
\begin{equation}
	\Nut = \frac{F_{\mathrm{T}}}{F_{\mathrm{cT}} } \qquad \mathrm{and} \qquad \Nus = \frac{F_{\mathrm{S}}}{F_{\mathrm{cS}}}
\end{equation}
with $F_{\mathrm{T}}$ being the total heat flux, $F_{\mathrm{cT}}$ the flux 
that would be transported if the temperature profile was linear between the bottom and the top, $F_{\mathrm{S}}$ the measured saline flux and $F_{\mathrm{cS}}$ the saline flux that would be transported if the concentration 
profile was linear between the bottom and the top. 

\begin{figure}
	\begin{minipage}[]{0.49\textwidth}
		\includegraphics[width=\textwidth]{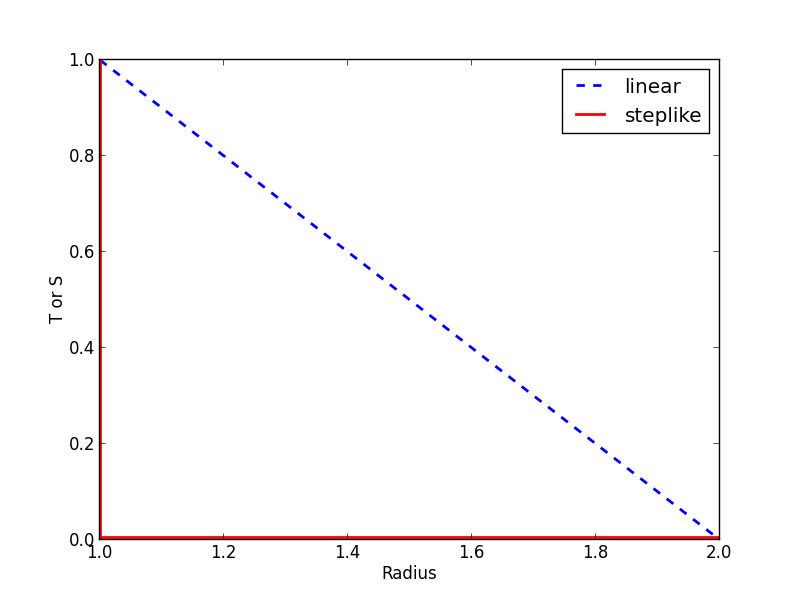}
	\end{minipage}
	\begin{minipage}[]{0.49\textwidth}
		\includegraphics[width=\textwidth]{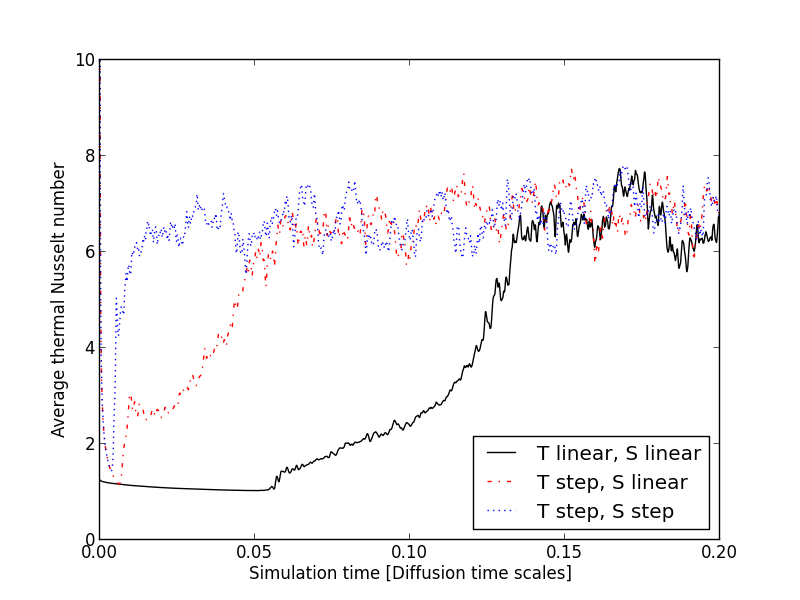}
	\end{minipage}
	\caption{Left-hand side: step-like and linear initial distributions of temperature or salinity as discussed in the text. 
	Right-hand side: average thermal Nusselt number vs. simulation time in thermal diffusion time scales for three different initial conditions for temperature (T) and salinity (S): a step in both T and S, 
	a step in T and a linear distribution in S and linear distributions for both T and S. 
	It can easily be seen that each initial condition leads to the same asymptotic range of values for the average thermal Nusselt number. The time that it takes to reach this
	asymptotic value differs, however. $\Rat = 10^7, \Pran=1, \Le=0.1$}
	\label{fig:ic_vgl}
\end{figure}

The result is shown in figure \ref{fig:ic_vgl}. %
Each initial condition leads to the same asymptotic range of values for the average thermal Nusselt number which means that the physical state after relaxation is the same.  
Only the time it takes to reach this state differs.
For ``T step, S step'', plumes immediately reached the upper boundary of the shell without any layering in between. But as we wanted to investigate
the effects of rotation on layering, this initial condition was no viable option for us.

To reduce computational costs, we did not chose ``T linear, S linear''. 
This left us with the initial condition of a step in the temperature field and a
linear distribution of salinity which offered a good compromise between observing layering and keeping simulation time within affordable limits.

\subsubsection{Parameter regime}
We investigated three different parameter regimes. The main simulations were run with $\Pran=1, \Le=0.1$ and $\Rat= 10^7$. To be able to reach the layered convective state, $R_\rho$ has to be sufficiently 
small. There exist two upper bounds for the maximum value of $R_\rho$, for which layer formation occurs. The one given by linear stability analysis is $R_{\rho,\mathrm{max}} = (1+\Pran) / (\Le + \Pran)$ which gives
$R_{\rho,\mathrm{max,lin}} = 1.8$ in our case. For larger values, the flow would be damped by viscous friction.
The other one is given by the model of \citet{spruit_theory_2013} (figure 3 therein) and is $R_{\rho,\mathrm{max,Spuit}} \approx 1.6$ with our parameters and provides an upper limit for the subcritical instability that triggers
layer formation.

Additionally, to avoid the simple case of convective mixing due to an unstable stratification in the sense of \citet{ledoux_1947}, $R_{\rho}$ should be larger than $1$. 
In simulations without rotation an increase of $R_\rho$ has a stabilising effect
on the flow \citep{zaussinger_diss, zaussinger_scn_2013,Rosenblum2011} and different regimes as a function of $R_{\rho}$ are established (figure 3 and chapter 3.2 in \citet{zaussinger_scn_2013} and the schematic 
illustration in figure 1 of \citet{mirouh2012}). We have investigated if the same applies when rotation is present and 
have chosen $R_\rho=1.3$ for our main simulations and $R_\rho=1.5$ for comparison runs. 
In another set of comparison runs we reduced the Prandtl number to $0.5$ to get a hint of the effects of viscosity on the rotational constraints. \\

A commonly used dimensionless number used to measure the
respective importance of buoyancy and rotation on a system is the Rossby number
\begin{equation}
	\Ro = \frac{V}{2 \Omega L \, \mathrm{sin}(\Lambda)}
	\label{eq:rossby}
\end{equation}
where $V$ is the characteristic flow speed and $\Lambda$ is the colatitude.

This can be written as
\begin{equation}
 Ro = \frac{1}{\mathrm{sin}(\Lambda)}\sqrt{\frac{Ra}{Pr Ta}}.
 \label{eq:ro_taylor}
\end{equation}

In order for a motion to be significantly influenced by rotation the Rossby number must be of
order one or less. 
We have chosen a range of Rossby numbers
near unity: $0.1, 0.3, 0.5, 1, 3, 4, 10$ and the case without rotation. Since the used code uses the Taylor number as an input parameter, we rewrite (\ref{eq:ro_taylor}) as 
\begin{equation}
 Ta = \frac{Ra}{(Ro \cdot \mathrm{sin}(\Lambda))^2 Pr }.
\end{equation} 
A point to note is that for the same Rossby number, we have
different Taylor numbers when the Prandtl number varies. This is important because we ran simulations with Prandtl numbers $1$ and $0.5$. Accordingly, the Taylor numbers of the simulations with $\Pr = 0.5$ had to be
doubled so that the Rossby number was the same.

The corresponding Taylor numbers for Prandtl numbers 0.5 and 1 at different colatitudes are shown in table \ref{tab:taylor_rossby}.

\begin{table}
\begin{center}
\begin{tabular}{ccccccc}
	\hline
\multicolumn{3}{c}{Taylor number at} & \multicolumn{4}{c}{Rossby Number at colatitude} \\
\multicolumn{1}{l}{$Pr=1$} & \multicolumn{1}{l}{$Pr=0.5$} & &\multicolumn{1}{c}{$\Lambda = \pi/2$} & \multicolumn{1}{c}{$\Lambda = \pi/3$} & \multicolumn{1}{c}{$\Lambda=\pi/4$} &
\multicolumn{1}{c}{$\Lambda=\pi/6$} \\ \hline \hline
$1.00\cdot 10^9$ & $2.00\cdot 10^9$ & & 0.10 & 0.12 & 0.14 & 0.20 \\ 
$1.11\cdot 10^8$ & $2.22\cdot 10^8$ & & 0.30 & 0.35 & 0.42 & 0.60 \\ 
$4.00\cdot 10^7$ & $8.00\cdot 10^7$ & & 0.50 & 0.58 & 0.71 & 1.00 \\ 
$1.00\cdot 10^7$ & $2.00\cdot 10^7$ & & 1.00 & 1.15 & 1.41 & 2.00 \\ 
$1.11\cdot 10^6$ & $2.22\cdot 10^6$ & & 3.00 & 3.46 & 4.24 & 6.00 \\ 
$4.00\cdot 10^5$ & $8.00\cdot 10^5$ & & 5.00 & 5.77 & 7.07 & 10.00 \\ 
$1.00\cdot 10^5$ & $2.00\cdot 10^5$ & & 10.00 & 11.55 & 14.14 & 20.00 \\ 
$0$ & $0$ & & $\infty$ &$\infty$  & $\infty$  & $\infty$  \\ \hline
\end{tabular}
\end{center}
\caption{Rossby numbers and corresponding Taylor numbers for Prandtl numbers $1$ and $0.5$}
\label{tab:taylor_rossby}
\end{table}

\section{Numerical implementation}\label{sec:numerical_implementation}
The numerical solution of the Navier-Stokes equations in the spherical shell is based on a spectral method. The radial direction is discretised on Chebyshev nodes ($k$), which brings desired clustering at the boundaries. However,
the variables are expanded in spherical harmonics along the meridional ($l$) and the zonal ($m$) direction. While the linear parts of the equations are solved in the spectral space, the non-linear ones are calculated in real space. The
time integration is based on a modified Runge-Kutta scheme of second order, where the diffusive terms are treated implicitly. This numerical method is fast and accurate especially for double-diffusive flows with high Prandtl
number.  Another advantage of the method is the uncoupled resolution of the vector fields and the scalar fields. This means that the expansion of each variable can be treated independently. Mainly low Mach number flows benefit
from this feature.  The velocity field is discretised on $k=71$ nodes in radial direction, $l=71$ modes in meridional and $m=71$ modes in zonal direction, respectively, for all simulations. The expansion of the
temperature and the solute equations use $k=71$ radial nodes and $l=m=71$ modes. The time step is set to $\tau=2 \cdot 10^{-6}$ thermal diffusion time scales.

We refer to \citet{hollerbach_2000} for a detailed description of the method that was presented here briefly. In contrast to the
original version of the code, the set of equations was extended for solute transport. This introduces the Lewis number $Le=\kappa_S/\kappa_T$ and the stability ratio $R_{\rho}=Ra_S/Ra_T$ as new input parameters.

\section{Results}\label{sec:results}
\subsection{The non-rotating reference run: $\Pran=1, R_{\rho}=1.3, \Le=0.1, Ra_T=10^7,\Ta=0$  }\label{sec:nonrotating}
\begin{figure}
	\begin{center}
		\includegraphics[width=\textwidth]{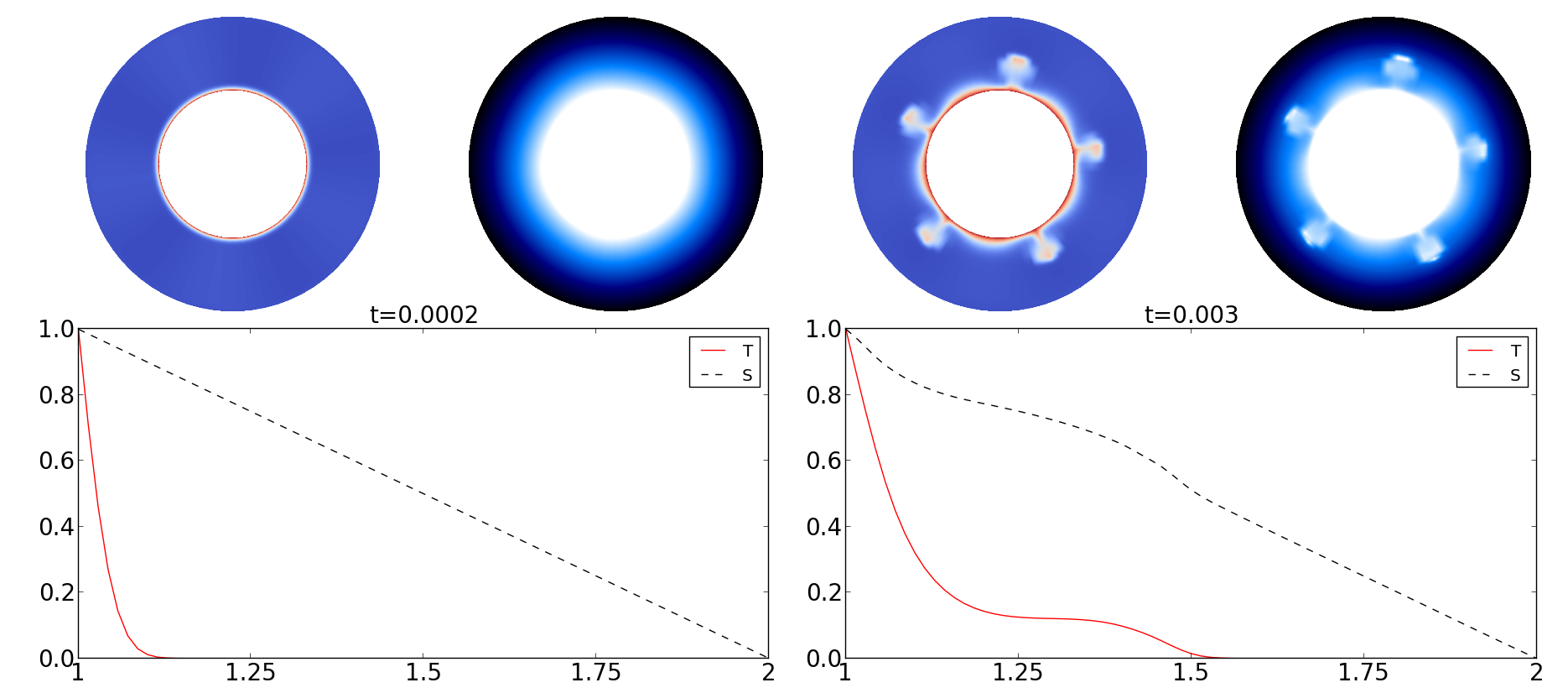}
	\end{center}
	\begin{center}
		\includegraphics[width=\textwidth]{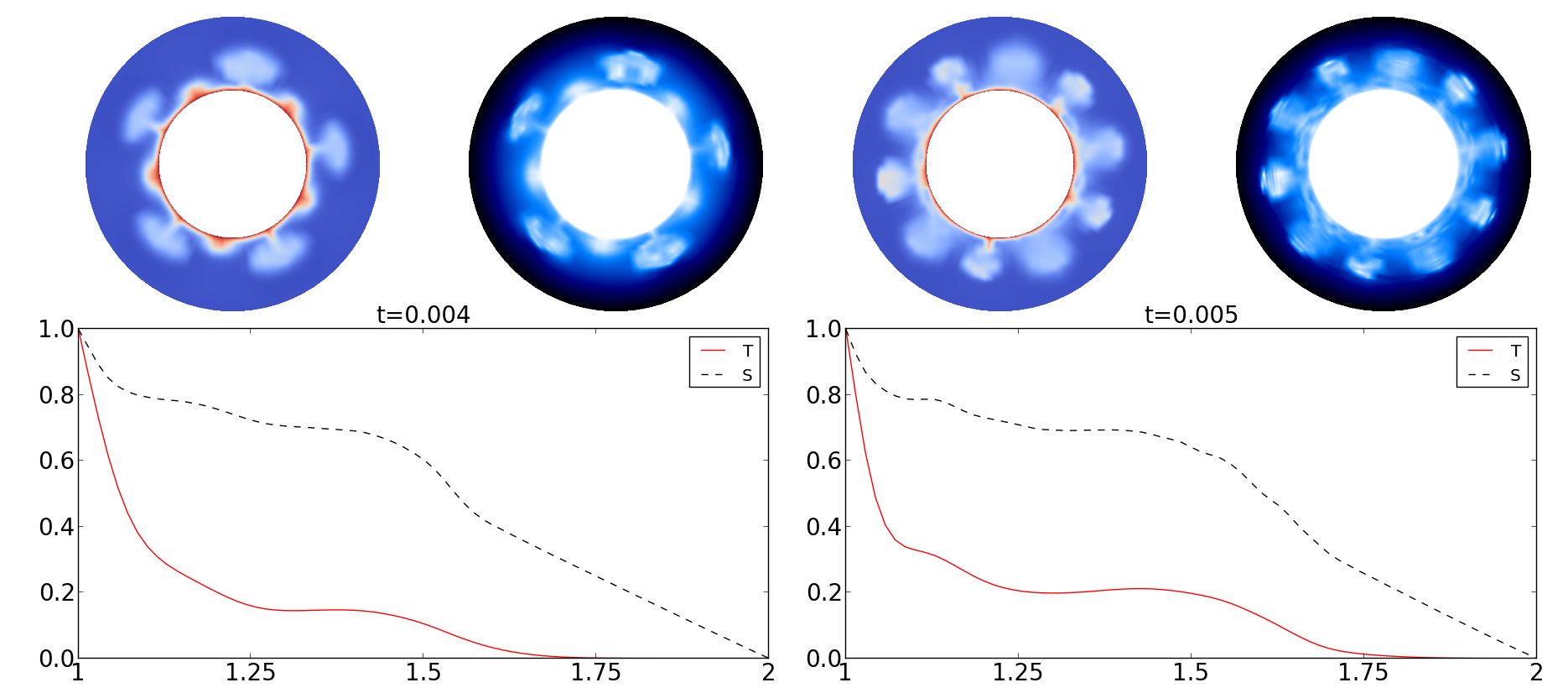}
	\end{center}
	\begin{center}
		\includegraphics[width=\textwidth]{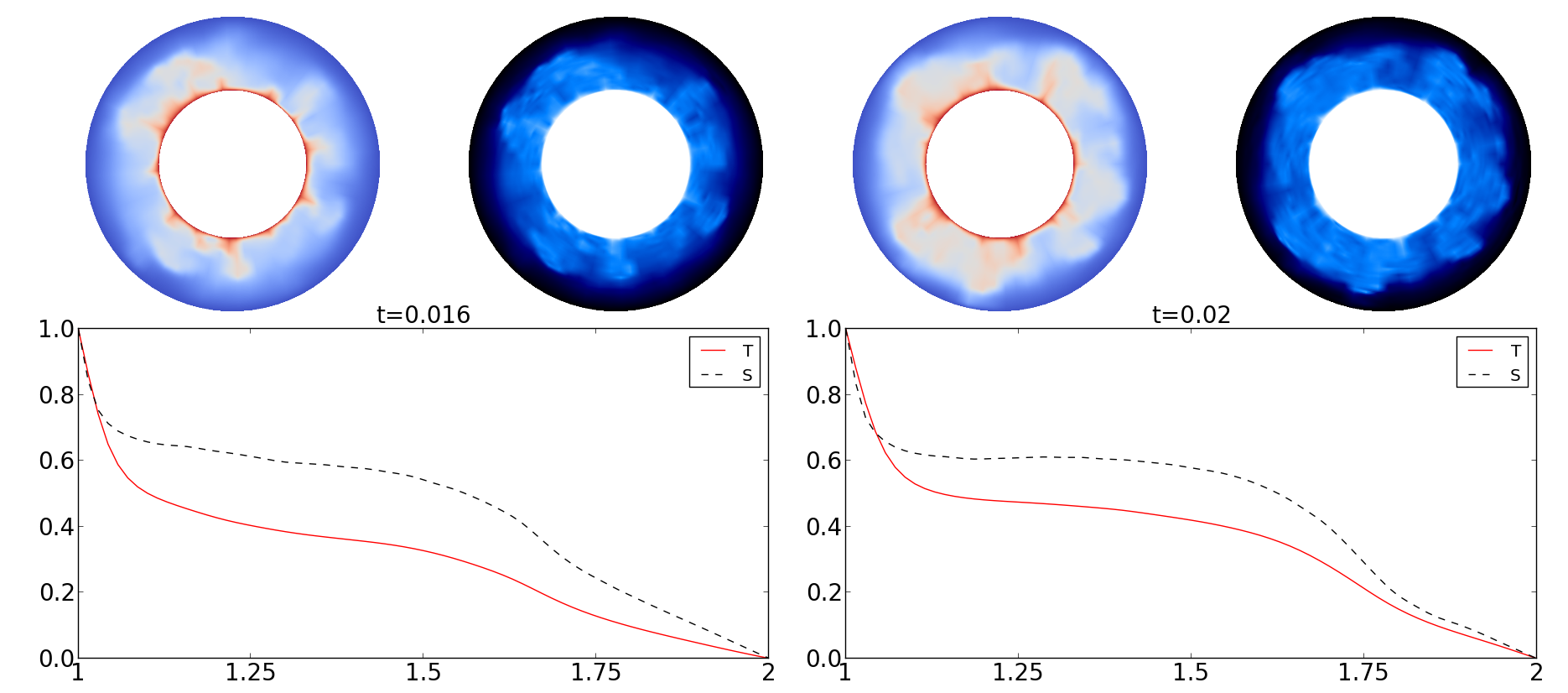}
	\end{center}
	\caption{Temporal evolution of the temperature field (left above each plot) and the salinity field (right above each plot) for $\Ta=0$ and plots of averaged potential 
	temperature (T) and salinity (S) in the equatorial plane vs. radius. Note: the snapshots are not equidistant in time.}
	\label{fig:taylor0overview}
\end{figure}
\begin{figure}
	\begin{center}
		\includegraphics[width=\textwidth]{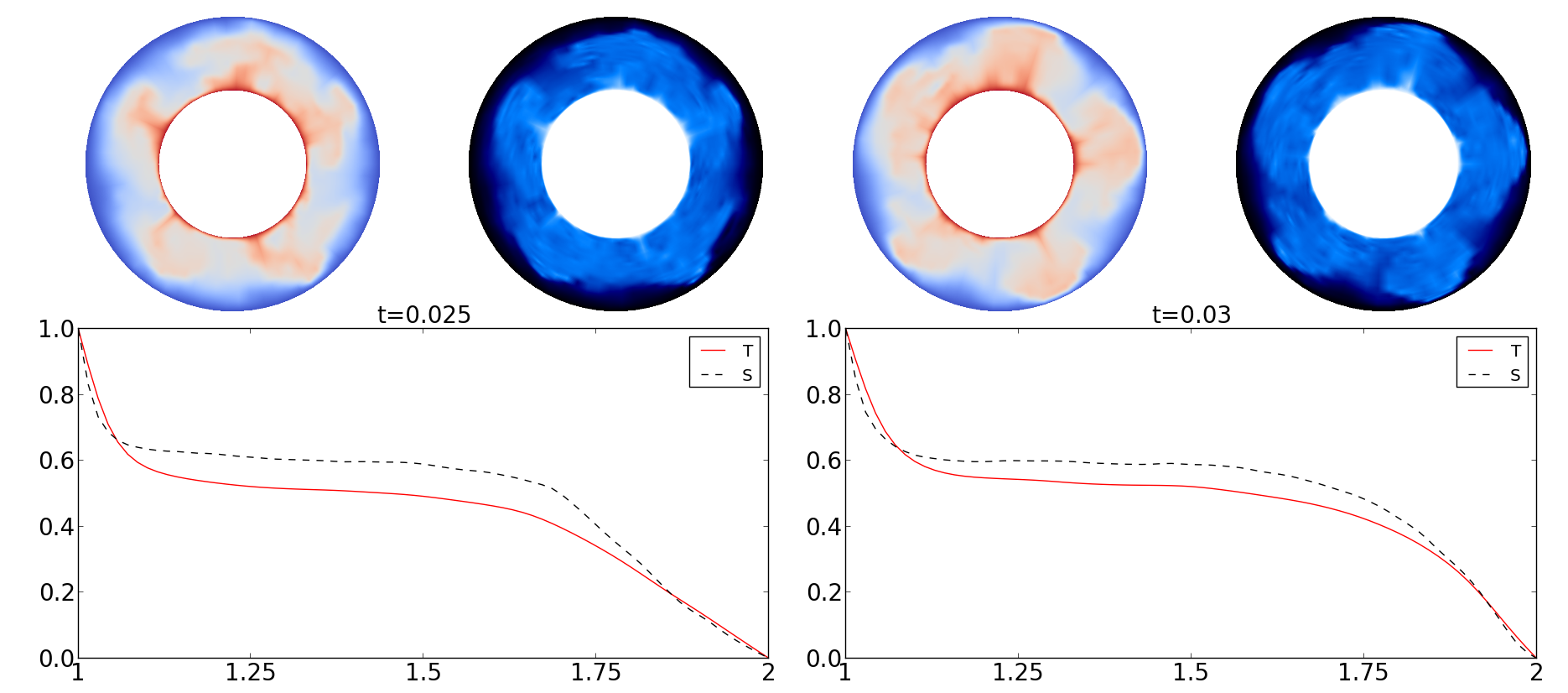}
	\end{center}
	\begin{center}
		\includegraphics[width=\textwidth]{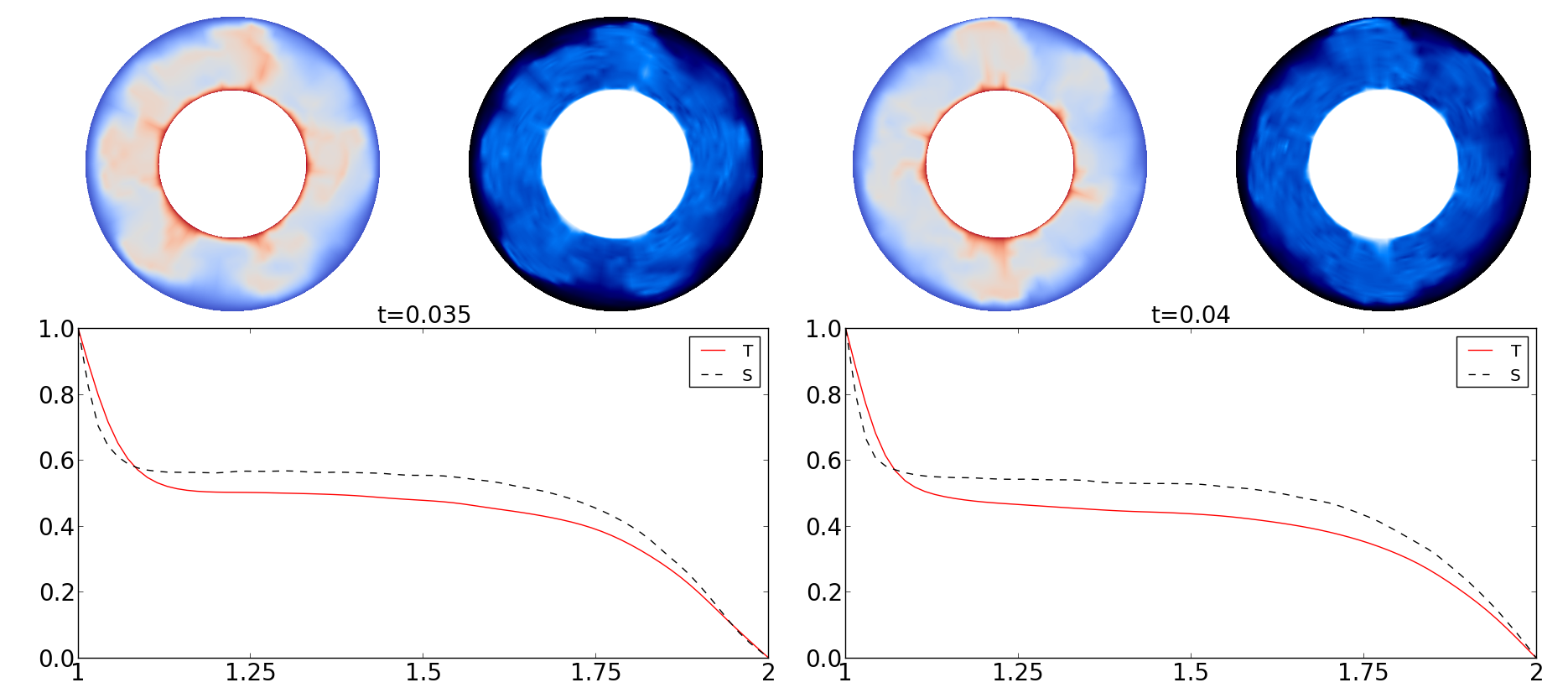}
	\end{center}
	\begin{center}
		\includegraphics[width=\textwidth]{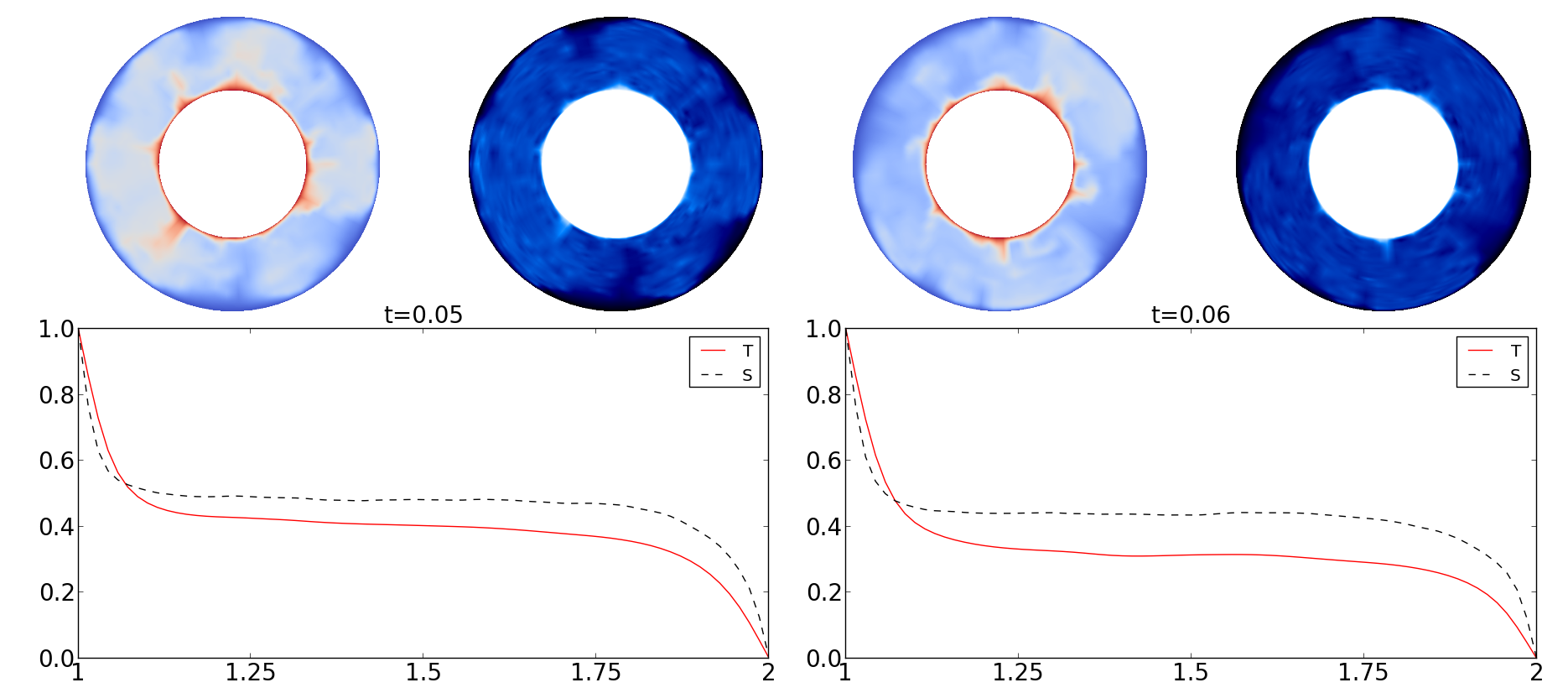}
	\end{center}
	\caption{Temporal evolution of the temperature field (left above each plot) and the salinity field (right above each plot) for $\Ta=0$ and plots of averaged potential temperature (T) and salinity (S) 
	in the equatorial plane vs. radius. Note: the snapshots are not equidistant in time.}
	\label{fig:taylor0overview2}
\end{figure}

\begin{figure}
	\begin{center}
		\includegraphics[width=\textwidth]{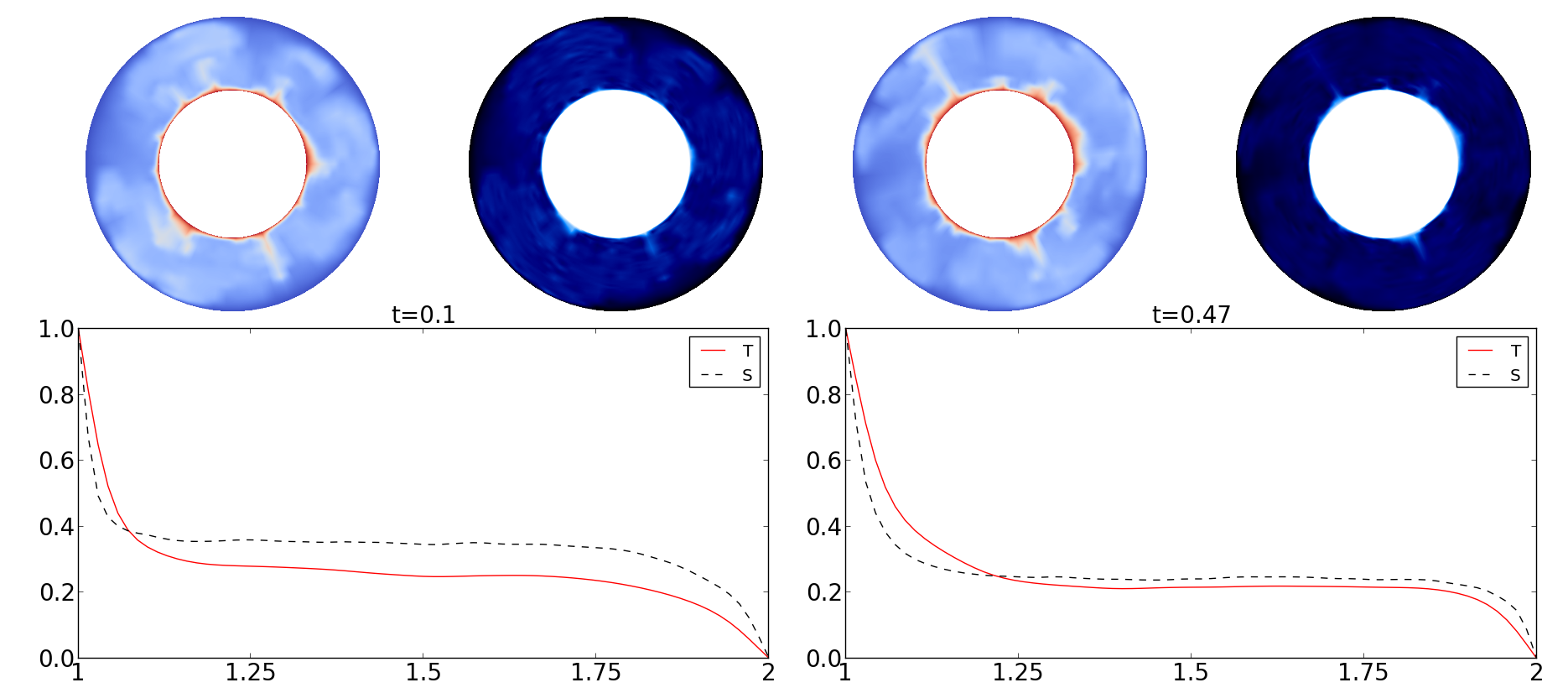}
	\end{center}
	\caption{Temporal evolution of the temperature field (left above each plot) and the salinity field (right above each plot) for $\Ta=0$ and plots of averaged potential temperature (T) and salinity (S) 
	in the equatorial plane vs. radius. Note: the snapshots are not equidistant in time.}
	\label{fig:taylor0overview3}
\end{figure}

First, we take a look at the temporal evolution of the reference run with parameters $\Pran=1$ and stability parameter $R_{\rho}=1.3$.

In order to validate the reference simulation we compare the numerical outcome with theoretical results. The convective flux parameterised as Nusselt number $Nu_T$ follows typically a power law in the form of $Nu_T=a Ra_T^b$,
whereas $0.1< a < 0.3$ is a constant factor and $b\approx 2/7$. However, dozens of different power laws have been found until now, which makes it nearly impossible to compare two simulations exactly. The $2/7$ power law seems to
be valid for most applications, which gives $a=0.19/\pi$ for our simulations. This is in good agreement with \citet{Castaing1989} and \citet{Kerr1996}. To check the saline flux, we come back to the linear stability analysis, e.g.
\citet{huppert1976}, which gives $Nu_S=Le^{-1/2} Nu_T$ for $Nu_T>>1$. A correction for $Nu_T=\mathcal O(1)$ was considered by \citet{spruit_theory_2013} and tested by \citet{zaussinger_scn_2013}, 
\begin{equation}
Nu_S-1=q Le^{-1/2}/R_{\rho} \, (Nu_T-1),
\label{NuS-NuT}
\end{equation}
where $q\approx1$ is a fitting parameter. Our reference simulation gives mean values of $Nu_T=5.77$ and $Nu_S=11.62$, thus (\ref{NuS-NuT}) seems to fit very well for $q=0.915$. 

To get an overview, the temporal evolution of the temperature and salinity fields in a semiconvective setup in a non-rotating spherical shell is shown in figure \ref{fig:taylor0overview}. 
\begin{figure}
	\begin{minipage}{0.5\linewidth}
	\begin{center}
		\includegraphics[width=\textwidth]{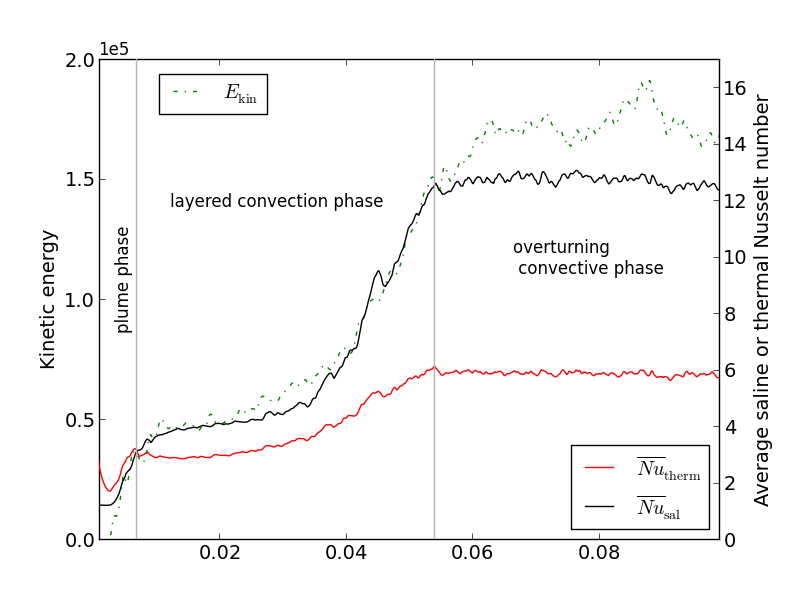}
	\end{center}
	\end{minipage}
	\begin{minipage}{0.5\linewidth}
	\begin{center}
		\includegraphics[width=\textwidth]{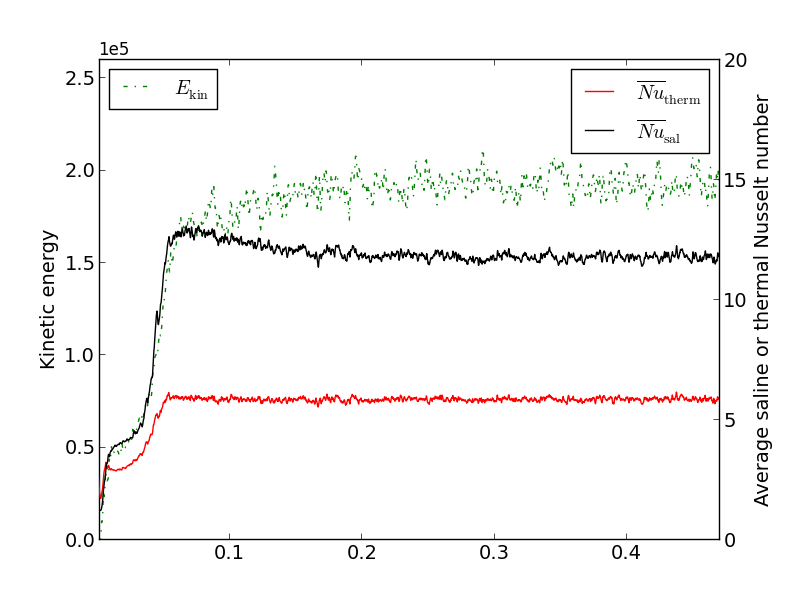}
	\end{center}
	\end{minipage}
	\caption{Kinetic energy (highest curve), average saline (middle curve) and average thermal (lowest) Nusselt numbers vs. simulation time for $\Ta=0$ from $t=0$ to $t=0.01$ (left) and from $t=0$ to $t=0.47$ (right)}
	\label{fig:thrice_taylor0}
\end{figure}
After starting the simulation it takes some time until convection sets in. The plumes do not, however, rise to the top of the shell. Convection is restricted to a zone with a thickness $d$. 
$d$ increases as time passes until the top of the zone touches the upper boundary at $t\approx0.03$. At $t\approx0.05$ the zone not just touches the upper boundary at 
some points
but fills the whole simulation domain. At $t\approx0.47$, the whole region is thoroughly mixed. Also shown in figure
\ref{fig:taylor0overview} are plots of averaged potential temperature (T) and salinity (S) in the equatorial plane vs. radius. We can see the relaxation of the initial
conditions to the solution which consists of a plateau of constant temperature and salinity throughout the spherical shell. This plateau is typical for a convective region. It is interesting to see that at $t\approx0.02$ 
there is a similar plateau but is not as wide as the one at $t\approx0.47$, meaning that the convective overturning region is limited to a smaller region in the shell. This is 
is also visible in the temperature field
itself. Another point to note is the height of the plateau which decreases as time passes. Interestingly, the plateaus of saline and thermal composition reach their final height at different times. This is visible in 
figure \ref{fig:taylor0overview3}. At $t\approx0.1$ the plateau of $T$ is at $\approx 0.25$ while the plateau of $S$ is a bit less than $\approx 0.4$. At $t\approx 0.47$, the thermal plateau has only moved by a very modest amount to
$\approx 0.2$ while the saline plateau has dropped a comparatively large amount to $\approx 0.25$. 

This phenomenon can also be observed when looking at the convective
flux and kinetic energies contained in the system (figure \ref{fig:thrice_taylor0}). 

While both the thermal and saline Nusselt numbers reach their maximum at the same time 
$t\approx 0.054$, the thermal Nusselt number keeps that value. The saline Nusselt number, however, decreases until at $t\approx 0.2$ .
The Nusselt numbers also emphasise the distinction of the flow state into different phases through which the convective flow develops. We have the ``plume phase''
which is characterised by the first appearance and upward movement of the convective plumes. Once they break we enter the ``layered convection phase'' at $t\approx 0.006$. In this phase we can identify
two regions,
based on the slope of Nusselt numbers and kinetic energy: one region with a low slope, corresponding to the rising of the semiconvective layer to the upper boundary, and one region with a high slope, corresponding to the case that 
one semiconvective layer fills out all of the shell, but semiconvection is still taking place. At $t\approx 0.054$ , the thermal part of the semiconvective layer turns into a fully mixed layer. The saline 
part of the semiconvective layer, however, needs much longer than the thermal part to reach equilibrium. While this difference is not visible in figure \ref{fig:taylor0overview}, it is clearly 
visible in figure \ref{fig:thrice_taylor0} (b): while the thermal Nusselt number 
reaches its asymptotic limit at $t \approx 0.054$ , the saline Nusselt reaches the equilibrium state at $t \approx 0.2$ . This is about the same time at which the kinetic
energy reaches its limiting average value. This is understandable because the flux of both the solute and the temperature add to the kinetic energy, hence they are dependent on each other.
Looking at figure \ref{fig:taylor0overview2} we see that the convective layer has already reached the top boundary at $t\approx 0.06$ , so the constant thermal convective flux is no measure for
how long a semiconvective layer lasts until it reaches its final state. Thermal and saline processes clearly have different lifetimes and the longer timescale on which saline processes take place agrees well with $\Le < 1$.

Comparing figures \ref{fig:taylor0overview2} and \ref{fig:thrice_taylor0}, we observe that at $t \approx 0.03$, when the semiconvective layer reaches the top 
boundary, the slope of the saline Nusselt number starts to increase. This increase continues up to about $t \approx 0.055$ where it reaches a maximum. Looking at a later simulation time we see that this maximum is in fact
the global maximum of the saline Nusselt number. From there, it slowly decreases until it reaches an asymptotic limit at $t \approx 0.2$. 

In summary, we have observed a growing double-diffusive layer that fills out the whole volume at the end of the simulation. We will now study the effects of rotation on this process.

\subsection{The effects of rotation}\label{sec:rotation}
To help emphasising some effects, we split up the results into three time scales: 
$t=0-0.03$ is the time it takes the semiconvective layer to reach the upper
boundary in the non-rotating case (see figures \ref{fig:taylor0overview2} and \ref{fig:ratio_of_nusselts}). 
$t=0-0.1$ is a time scale on which it becomes clear that the time a simulation needs to run is higher than expected from dynamical (flow related) timescales and $t=0-1$ is the 
complete simulation time where global effects
are visible. Since the fields of temperature and salinity look alike in figures \ref{fig:taylor0overview} to \ref{fig:taylor0overview3}, we restrict ourselves to showing the temperature field from now on.
We start with a discussion of the initial development phase.

\subsubsection{The temporal evolution with rotation up to $t=0.03$ }
After $t=0.03$ the semiconvective layer has reached the upper boundary in the non-rotating case (see figures \ref{fig:taylor0overview} and \ref{fig:taylor0overview2}). 
The situation is very different in the rotating case as is shown
in figure \ref{fig:at_dts_003} and in the movie online:
\begin{figure}
	\begin{minipage}[]{\linewidth}
	\begin{center}
		\includegraphics[width=\textwidth]{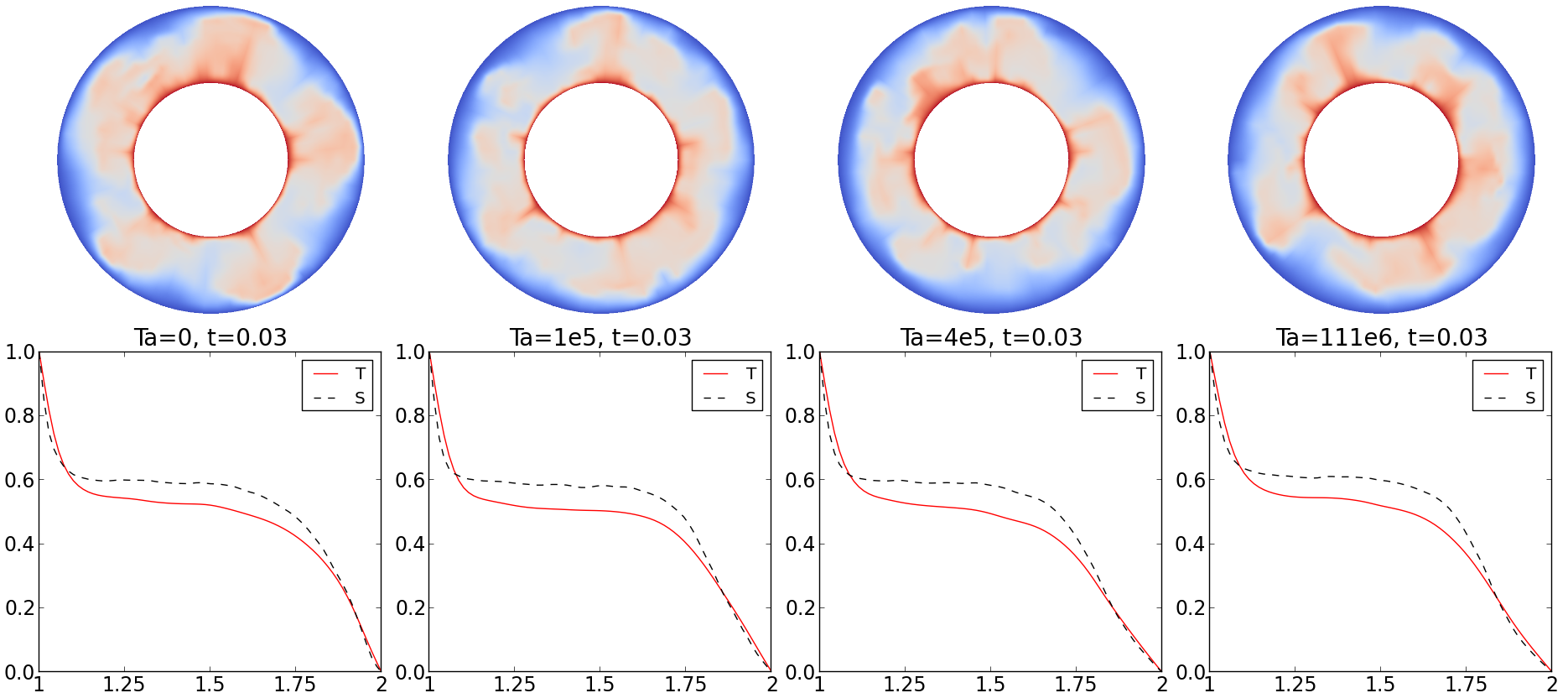}
	\end{center}
	\end{minipage}
	\begin{minipage}[ ]{\linewidth}
	\begin{center}
		\includegraphics[width=\textwidth]{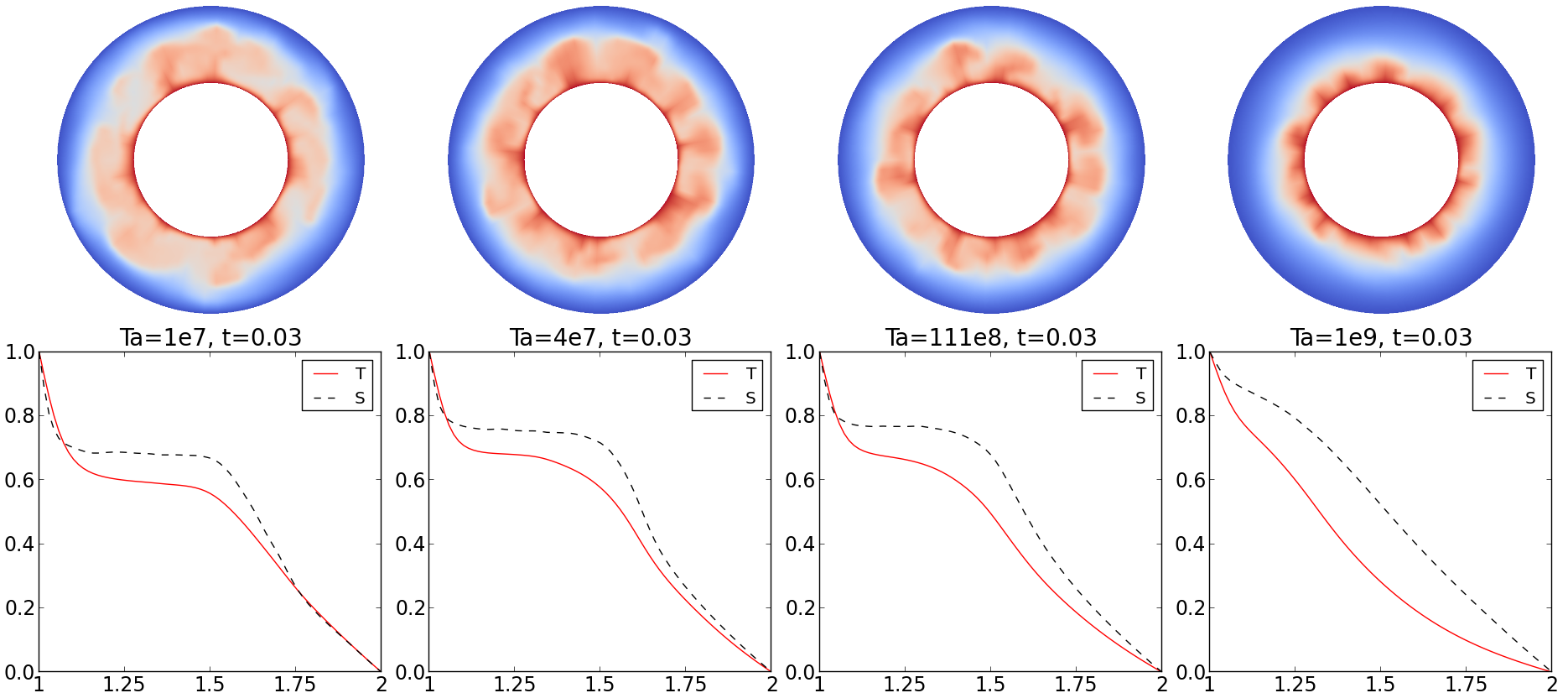}
	\end{center}
	\end{minipage}
	\caption{The temperature field at $t=0.03$ for different Taylor numbers and plots of averaged potential temperature (T) and salinity (S) in the equatorial plane vs. radius. See also the movie online for the temporal
	evolution of a few chosen Taylor numbers from $t=0$ up to $t=0.0564$.}
	\label{fig:at_dts_003}
\end{figure}
a moderate rotation rate already has a stabilising effect on semiconvection, similar to an increase of the salinity gradient.
At $\Ta=10^7$ this effect is very clearly visible. The semiconvective zone has a thickness of about 0.75.
Further out, temperature
and salinity diffuse out. At a certain critical Taylor number $\Tac$ convection is suppressed completely. A point to note is that if $\Ta < \Tac$ the onset of convection occurs practically simultaneously and is not influenced by 
the rate of rotation. This is shown in figure \ref{fig:dts003}: the Nusselt numbers start to increase at the same time $t\approx0.003$. It can also be seen in the movie supplementing the paper that convection starts at the same
time if $\Ta < \Tac$.

\begin{figure}
	\begin{minipage}[]{0.49\linewidth}
		\centering
		\includegraphics[width=\textwidth]{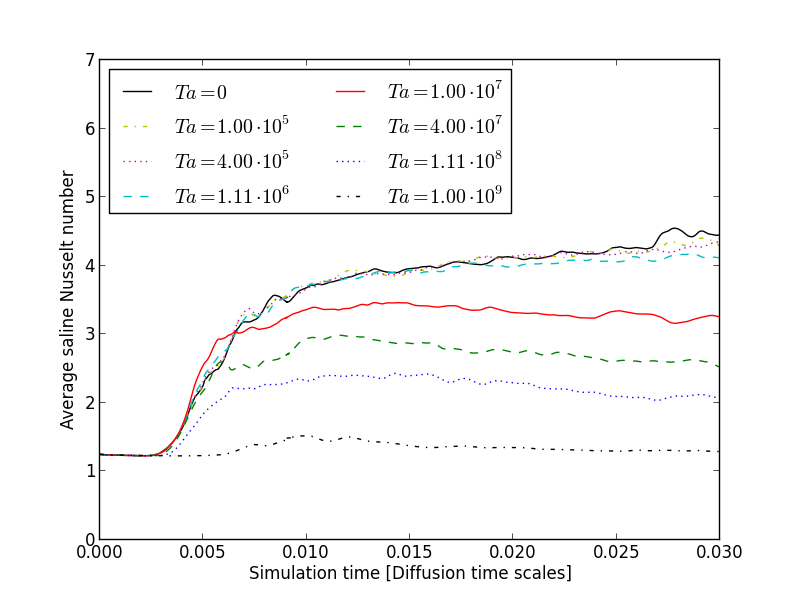}
	\end{minipage}
	\begin{minipage}[]{0.49\linewidth}
		\centering
		\includegraphics[width=\textwidth]{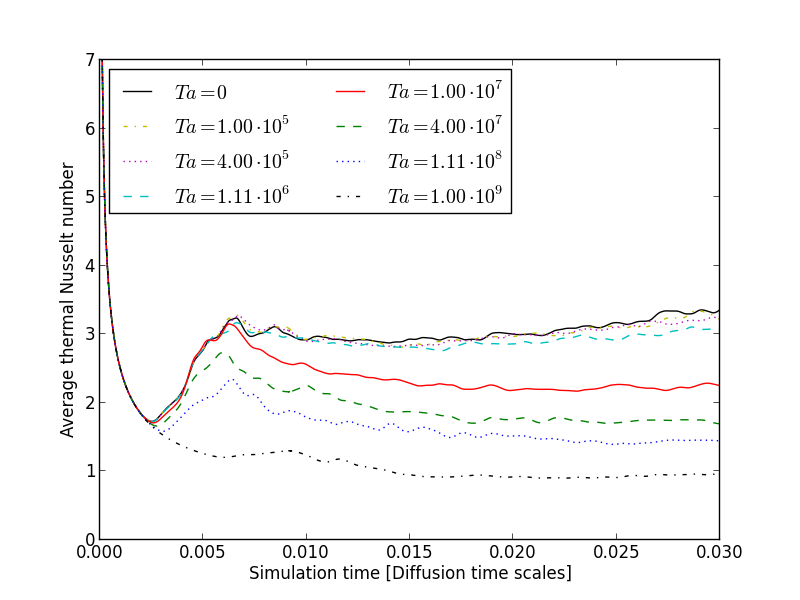}
	\end{minipage}
	\begin{minipage}[]{0.49\linewidth}
		\centering
		\includegraphics[width=\textwidth]{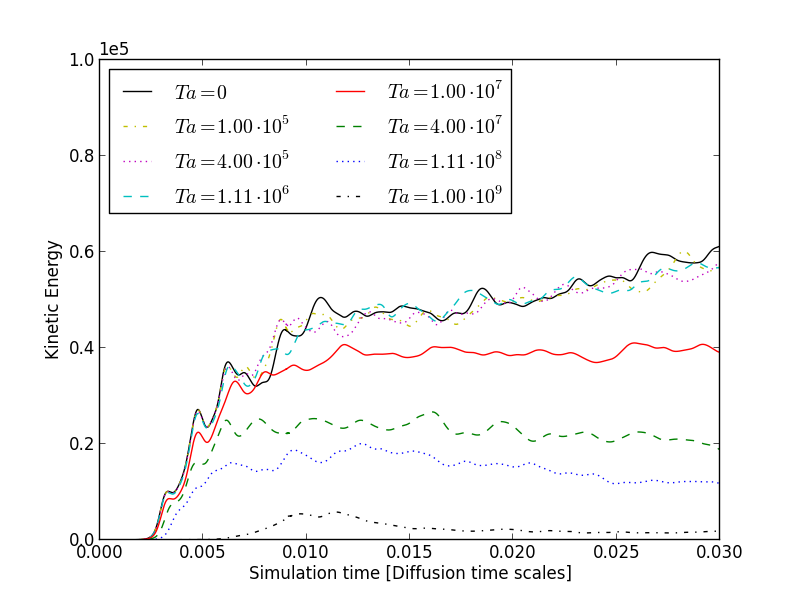}
	\end{minipage}
	\caption{Average saline and thermal Nusselt number and kinetic energy as function of simulation time for the specified Taylor numbers up to $t=0.03$ . Note the practically simultaneous onset of 
	 instability for $\Ta < 10^9$.}
	\label{fig:dts003}
\end{figure}

Figure \ref{fig:dts003} also shows the kinetic energy in the system and the average saline and thermal Nusselt numbers as a function of the simulation time up to $t=0.03$.
At higher rotation rates of $\Ta \geq 10^7$ both the Nusselt numbers and the kinetic energy have lower values than at lower rotation rates.

\subsubsection{The temporal evolution with rotation up to $t=0.1$ }

\begin{figure}
	\begin{minipage}[]{\linewidth}
	\begin{center}
		\includegraphics[width=\textwidth]{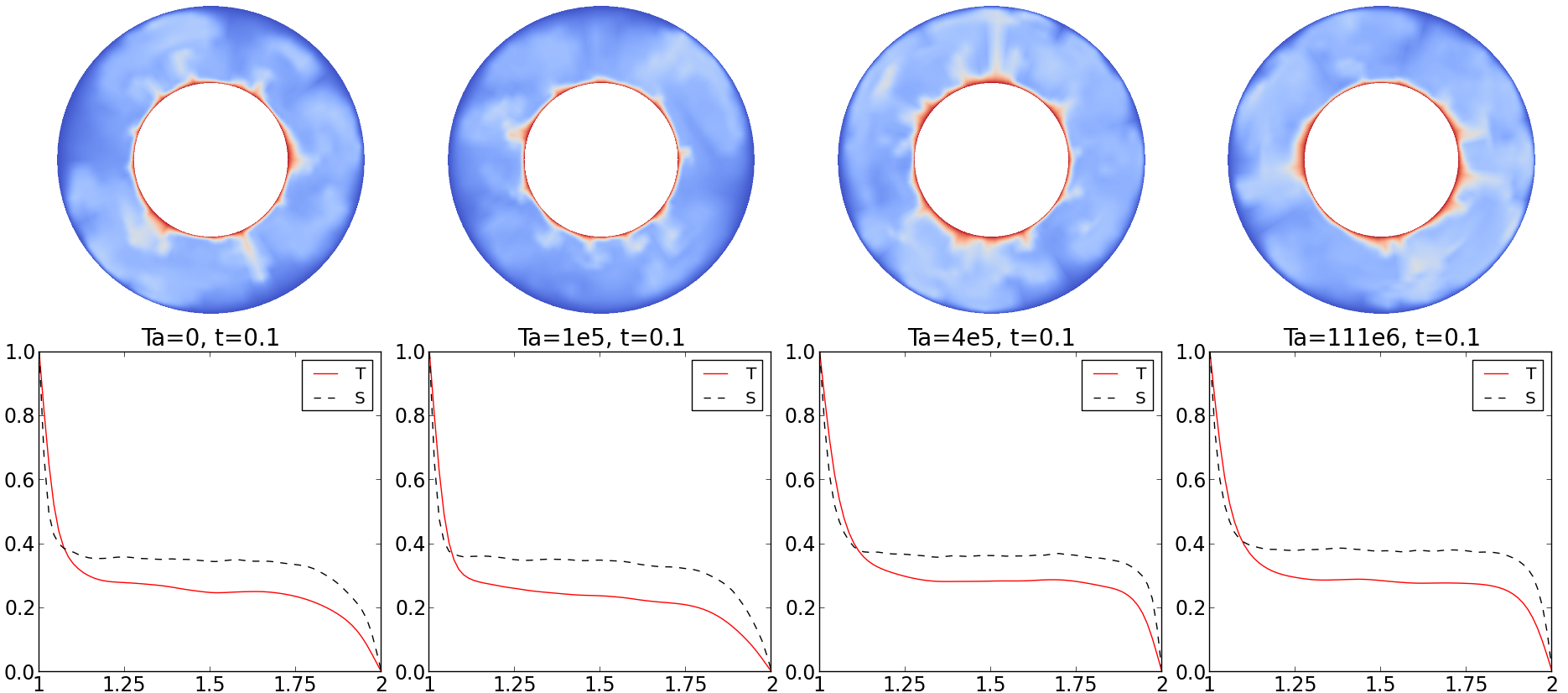}
	\end{center}
	\end{minipage}
	\begin{minipage}[]{\linewidth}
	\begin{center}
		\includegraphics[width=\textwidth]{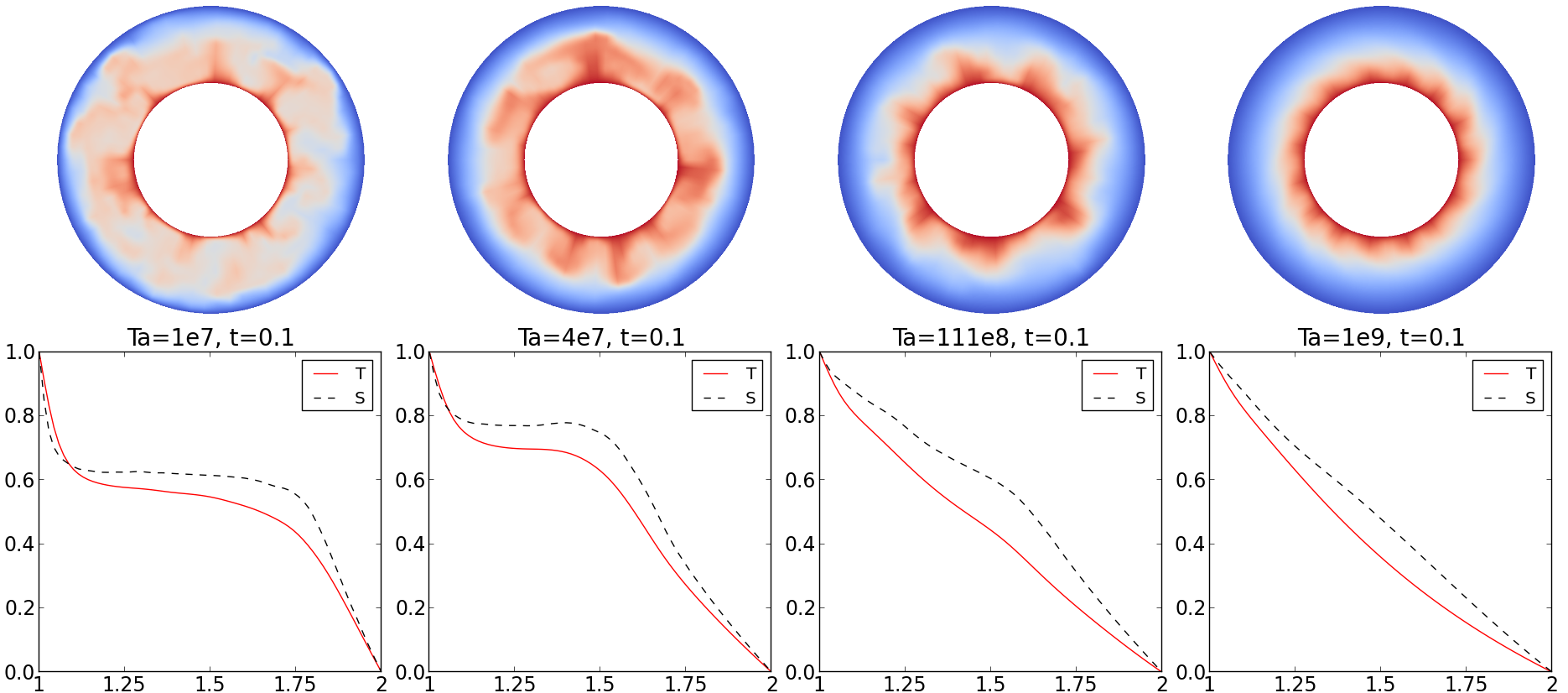}
	\end{center}
	\end{minipage}
	\caption{The temperature field at $t=0.1$ for different Taylor numbers 
	and plots of averaged potential temperature (T) and salinity (S) in the equatorial plane vs. radius.}
	\label{fig:at_dts_01}
\end{figure}

\begin{figure}
	\begin{minipage}[t]{0.49\linewidth}
		\centering
		\includegraphics[width=\textwidth]{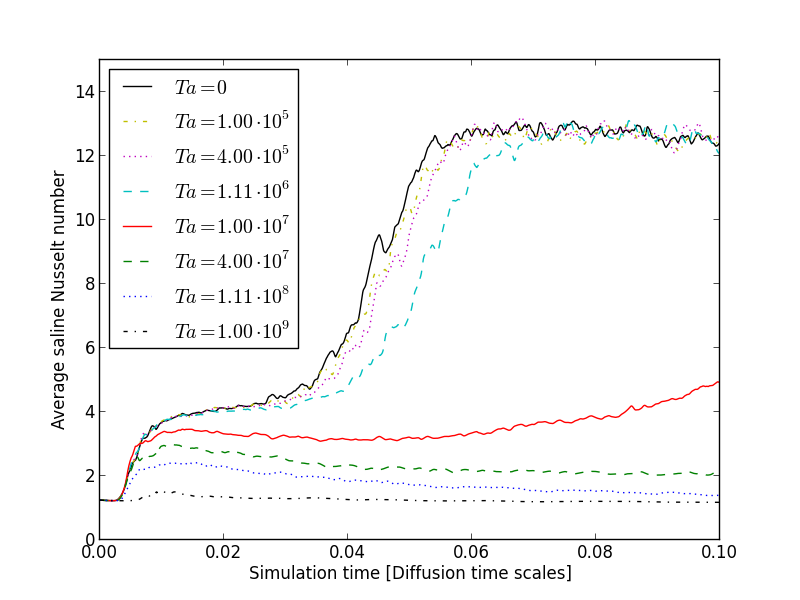}
	\end{minipage}
	\begin{minipage}[t]{0.49\linewidth}
		\centering
		\includegraphics[width=\textwidth]{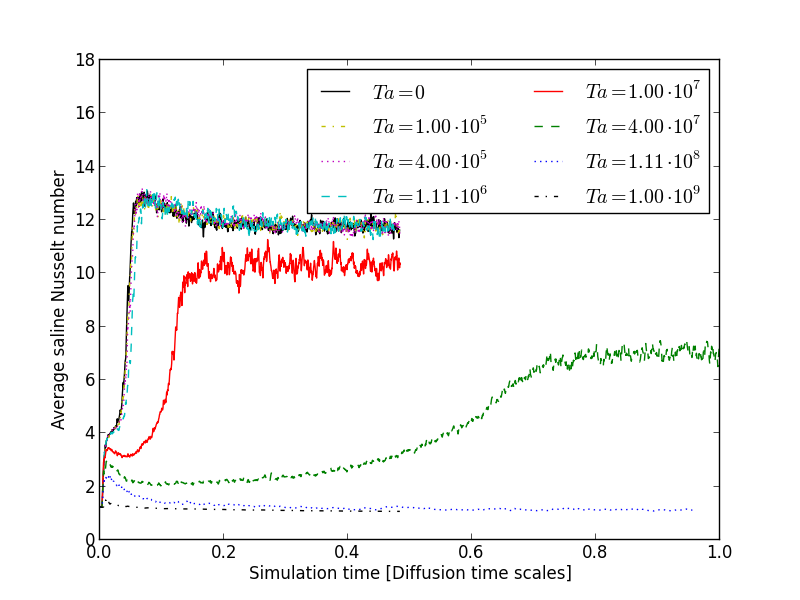}
	\end{minipage}
	\begin{minipage}[t]{0.49\linewidth}
		\centering
		\includegraphics[width=\textwidth]{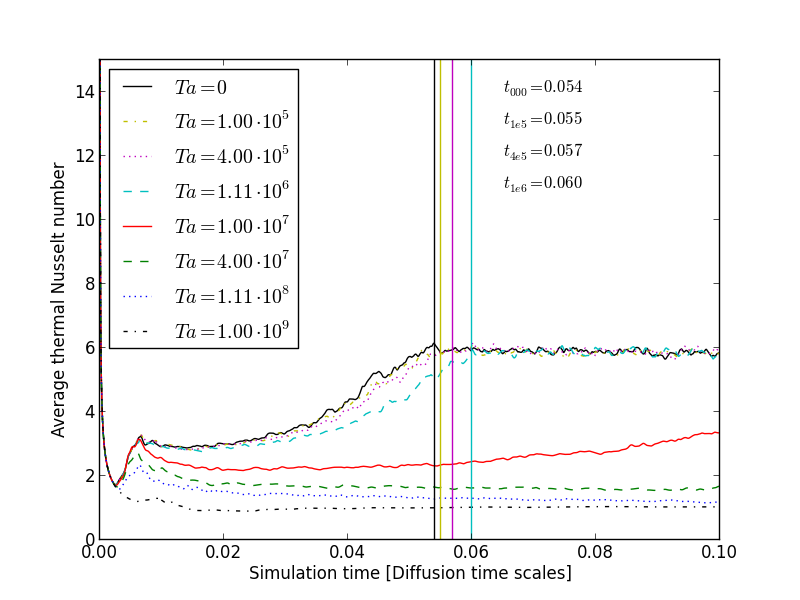}
	\end{minipage}
	\begin{minipage}[t]{0.49\linewidth}
		\centering
		\includegraphics[width=\textwidth]{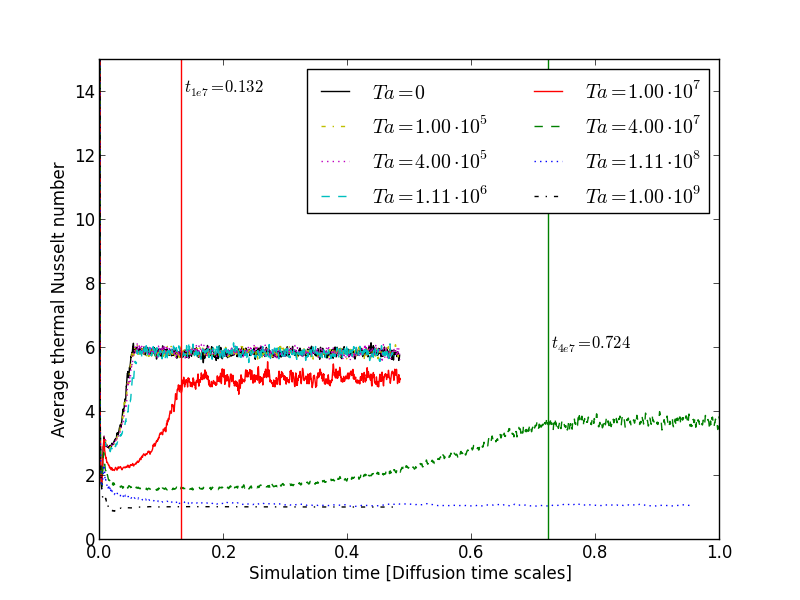}
	\end{minipage}
	\begin{minipage}[t]{0.49\linewidth}
		\centering
		\includegraphics[width=\textwidth]{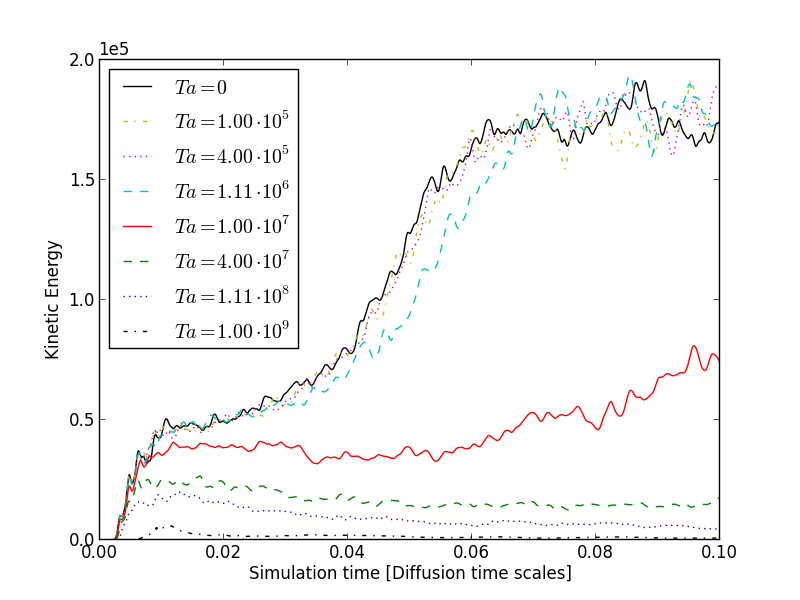}
	\end{minipage}
	\begin{minipage}[t]{0.49\linewidth}
		\centering
		\includegraphics[width=\textwidth]{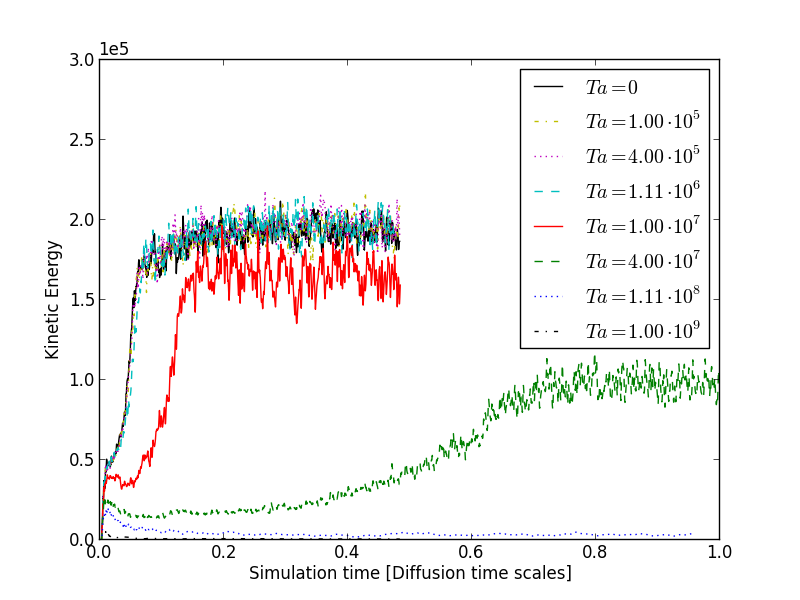}
	\end{minipage}
	\caption{Average saline and thermal Nusselt number and kinetic energy as function of simulation time for the specified Taylor numbers up to $t=0.1$ (left) and $t=1$ (right). The vertical bars represent the times when the
	average thermal Nusselt number reaches its statistically stable asymptotic state.}
	\label{fig:dts01udts1}
\end{figure}

We will now take a look at the real space temperature fields at $t=0.1$ for different rotation rates. These are shown in figure \ref{fig:at_dts_01} together with the averaged values of temperature 
and salinity.
For Taylor numbers $0, 10^5, 4 \cdot 10^5$ and $1.11 \cdot 10^6$ we can see that the whole spherical shell is thoroughly mixed and the convective plumes have reached the outer boundary. 
This results in a constant thermal Nusselt number as shown in figure \ref{fig:dts01udts1}. 
For Taylor numbers $10^7$ and $4 \cdot 10^7$ the simulation is still in the phase of semiconvective layering which is also visible in the real space pictures as well as in the plots of averaged $T$ and $S$. 
At $\Ta=1.11 \cdot 10^8$ no convective plateau is observable in the plot of averaged $T$ and $S$, so semiconvection already is seriously dampened by the high rate of rotation. This could 
correspond to the diffusive turbulent case which \citet[figure 3]{zaussinger_scn_2013} observe for simulations with high $R_\rho$, i.e. $R_\rho > R_{\rho,\mathrm{crit}}$ where $R_{\rho,\mathrm{crit}}$ is the maximum
value of $R_\rho$ for which layer formation can occur \citep{radko2003,spruit_theory_2013}.
At $\Ta=10^9$ there is no convection at all.\\

The left-hand column of figure \ref{fig:dts01udts1} shows the plots of kinetic energy and
average thermal and saline Nusselt numbers vs. time up to a simulation time of $t=0.1$. As in the temperature fields 
three distinctively different behaviours depending on the rotation rate are visible. 

The cases $\Ta=0, 10^5, 4 \cdot 10^5$ and
$1.11 \cdot 10^6$ are almost indistinguishable: the average thermal Nusselt number of these runs reaches an asymptotic limit of about $6$ like the non-rotating case..
Also, the four mentioned simulations have a similar temporal evolution of the convective flow: they all show an increase in energy and Nusselt numbers starting at $t \approx 0.02$  and an only slightly varying slope.
However, they do reach the 
asymptotic limit at different times. This will be investigated further when we take a look at the ratios of Nusselt numbers in chapter \ref{sss:ratio_nusselts_rotation}.

The simulation with $\Ta=10^7$ shows a distinct behaviour: energy and Nusselt number rise much slower than the simulations with a lower rotation rate but they are rising nonetheless. 

The simulations with
$\Ta=4 \cdot 10^7, 1.11 \cdot 10^8$ and $10^9$ show no increase of either kinetic energy nor Nusselt number. From looking at these plots alone, one could think that the simulations with these rotation rates 
will lead to a purely diffusive state and hence could be aborted. As we will see later, this would be a grave mistake because the case with $\Ta=4 \cdot 10^7$ is the most interesting one.

\subsubsection{The temporal evolution from $t=0.1$ to $t=1$ }
In order to get a clear picture of the global temporal evolution one has to look at the final equilibrium/unstable states which the systems take on after a very long simulation time. Figure \ref{fig:dts01udts1} shows the 
average saline and thermal Nusselt numbers and kinetic energy as a function of the simulation time for up to one full thermal diffusion time scale for all simulated Taylor numbers. Only after such a long time each simulation has 
reached its final state, as is indicated by the asymptotically constant Nusselt numbers.
Although Feudel et al. (2011) showed changes in convective patterns after 60 and more thermal time scales, these simulations are located in the laminar parameter space and the Nusselt number changes only in the second digit.
Once a simulation reaches the asymptotic limit of statistically constant Nusselt number, nothing changes in the convective 
state even after a longer simulation time. Letting the simulations
run any longer would provide no additional information. We regard every simulation which has reached this limit as finished. As already mentioned,
especially the simulation with $4 \cdot 10^7$ has attracted our attention. While it looked as if this rate of rotation had enough of a dampening effect on convection to suppress it completely 
in the left column of figure \ref{fig:dts01udts1} the picture is quite different if looking at the right column of figure \ref{fig:dts01udts1} or at the real temperature fields. Therefore, we need 
to point out that a sufficiently long simulation time is mandatory for this
kind of simulation in order to avoid any wrong conclusions. \\

\begin{figure}
	\begin{minipage}[]{\linewidth}
	\begin{center}
		\includegraphics[width=\textwidth]{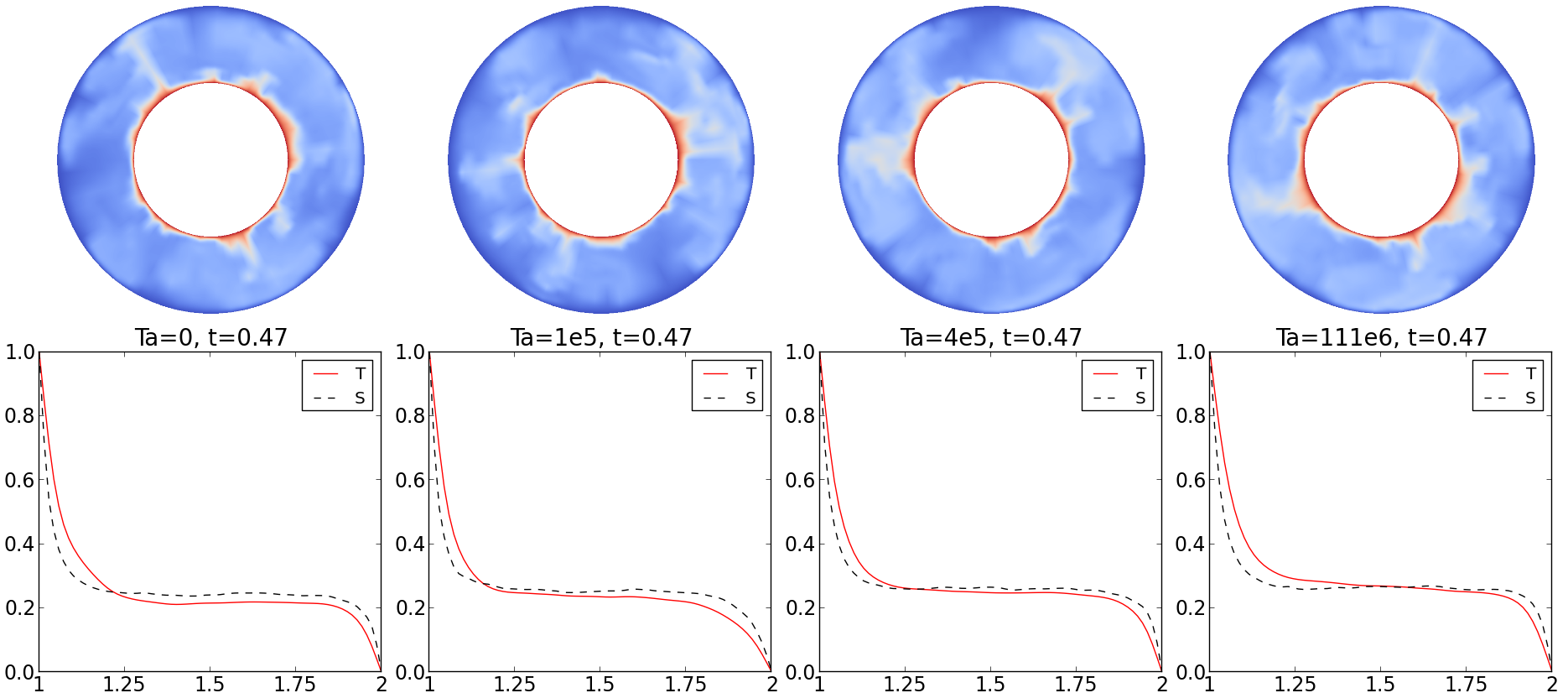}
	\end{center}
	\end{minipage}
	\begin{minipage}[]{\linewidth}
	\begin{center}
		\includegraphics[width=\textwidth]{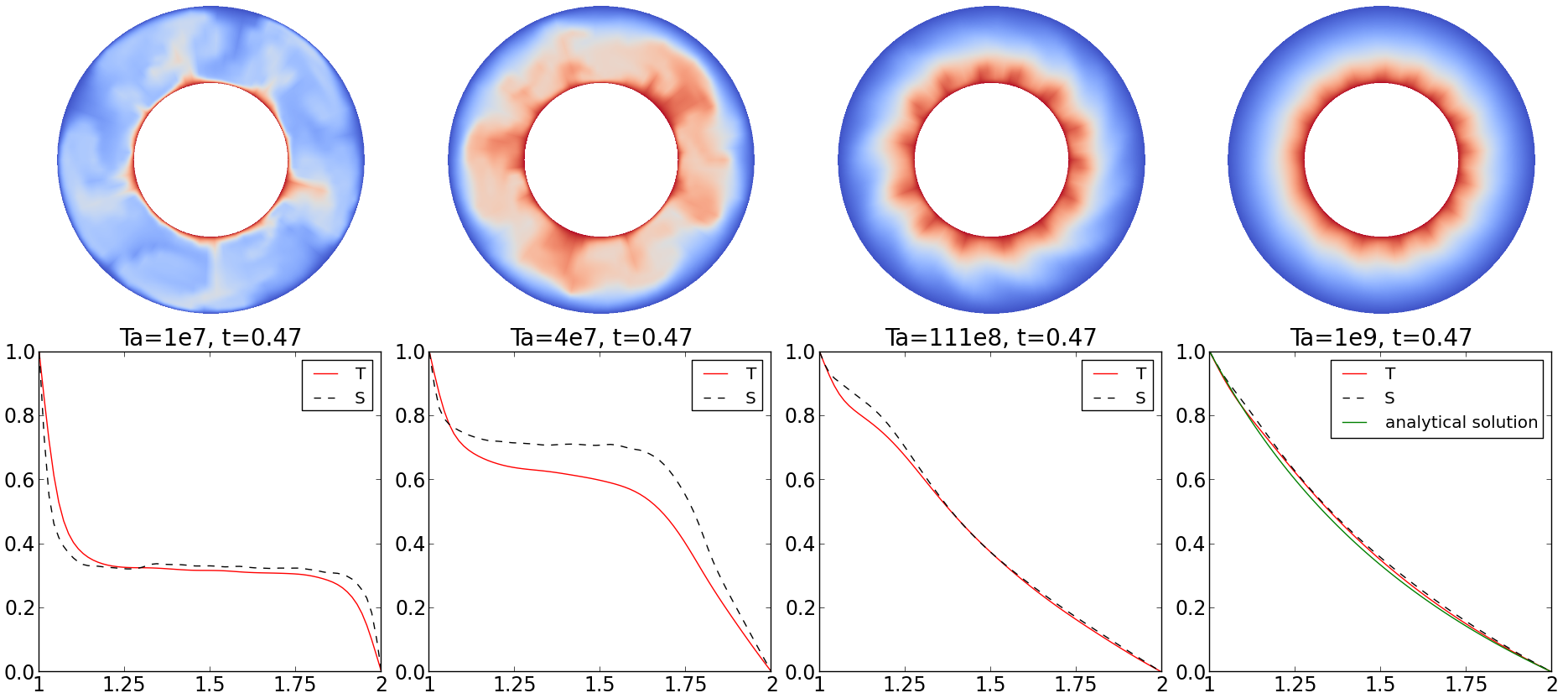}
	\end{center}
	\end{minipage}
	\caption{The temperature field at $t=0.47$ for different Taylor numbers 
	and plots of averaged potential temperature (T) and salinity (S) in the equatorial plane vs. radius. For $\Ta=10^9$, we plotted the analytical solution of (\ref{eq:A}) 
	as a means for comparison (see text).}
	\label{fig:at_dts_047}
\end{figure}

Having performed an analysis of the time evolution over one full thermal time scale lets us see very clearly that an increase of rotation slows down the temporal evolution and reduces the amount of the convective flux. 
The real space temperature fields after $t=0.47$  
are shown in figure \ref{fig:at_dts_047}. At that time, only the simulation with $\Ta=4\cdot10^7$ has not reached its final stable state yet. The ones with $\Ta=0 - 10^7$ have all reached the fully convective overturning state while
the ones with $\Ta=1.11 \cdot 10^8$ and $10^9$ show no convection at all and have diffused out. This is, in fact, close to the analytical solution of the 
diffusion problem:
when solving the diffusion equation 
\begin{equation}
	\frac{\partial T }{\partial t } = \kappa_T \nabla^2 T
	\label{eq:A}
\end{equation}
for the steady state in spherical coordinates and using as boundary conditions $T(R_1) = 1$ and $T(R_2) = 0$ and the fact that $R_2 = 2 R_1$, we obtain the solution 
\begin{equation}
	T(r) = \frac{2}{r} - 1 \quad \mathrm{for} \quad r \in [1,2].
	\label{eq:T_ana}
\end{equation}
This is plotted in figure \ref{fig:at_dts_047} for the case of $\Ta = 10^9$. We see that our numerical result converges nicely to the analytical one for the case of high rotation rates. 
Note that since $\kappa_T$ does not appear in (\ref{eq:T_ana}) --- basically because $\partial T / \partial t = 0$ in the steady state --- the solution for $S(r)$ is the same as for $T(r)$ thanks
to $S(R_1) = 1$ and $S(R_2) = 0$. This immediately follows from comparing (\ref{eq:ns3}) and (\ref{eq:ns4}) with (\ref{eq:A}), whence $S$ is governed by the same asymptotic laws and thus are found to converge to the same analytical 
solution in figure \ref{fig:at_dts_047}.

\subsection{Modifying $\Pran$ and $R_{\rho}$}\label{sec:modifying_pr_and_rrho}
A first conclusion we can draw from the previous observations is that rotation has a stabilising effect on the lifetime of semiconvective layers as a function of $\Ta$ 
with a variety of cases distinguished by the specific value of $\Ta$.

The next question we look into is, if the effects of rotation were similar
if we reduced the Prandtl number or increased the density ratio $R_\rho$. As we have seen, only rotation above a critical Taylor number has an effect on semiconvection. Because of this, we have neglected Taylor numbers
$0$ and $10^5$ in this chapter. This leaves us with six Taylor numbers which can be grouped into three categories based on the rate of rotation: low rotation rates ($\Ta=4\cdot10^5, 2.22\cdot10^6)$, 
medium rotation rates ($\Ta=4\cdot 10^6,2\cdot10^7$) and high rotation rates ($\Ta=1.11\cdot10^8, 2 \cdot 10^9$). We will have a look at each regime consecutively.

\subsection{The effect of different $\Pran$ and $R_{\rho}$ at low rotation rates}

\begin{figure}
	\centering
	\begin{minipage}[t]{.49\linewidth}
		\centering
		\includegraphics[width=1.0\textwidth]{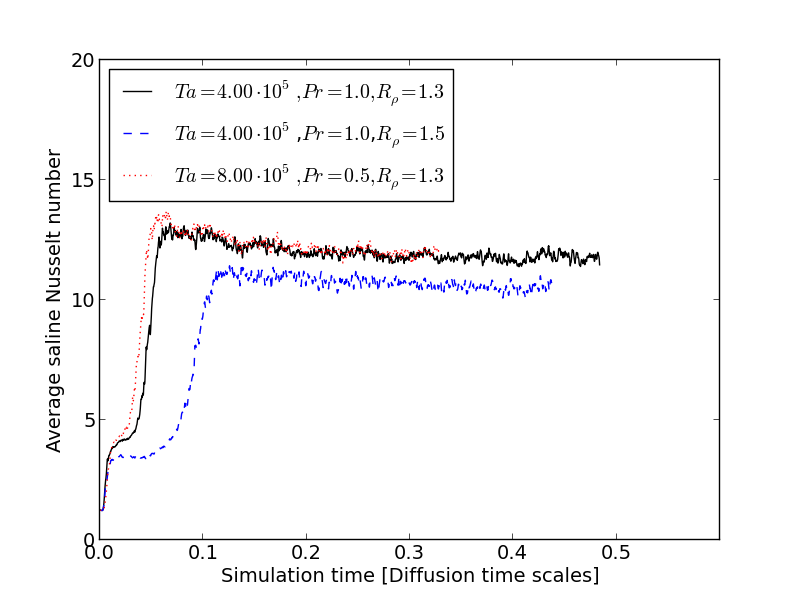}
	\end{minipage}%
	\begin{minipage}[t]{.49\linewidth}
		\centering
		\includegraphics[width=1.0\textwidth]{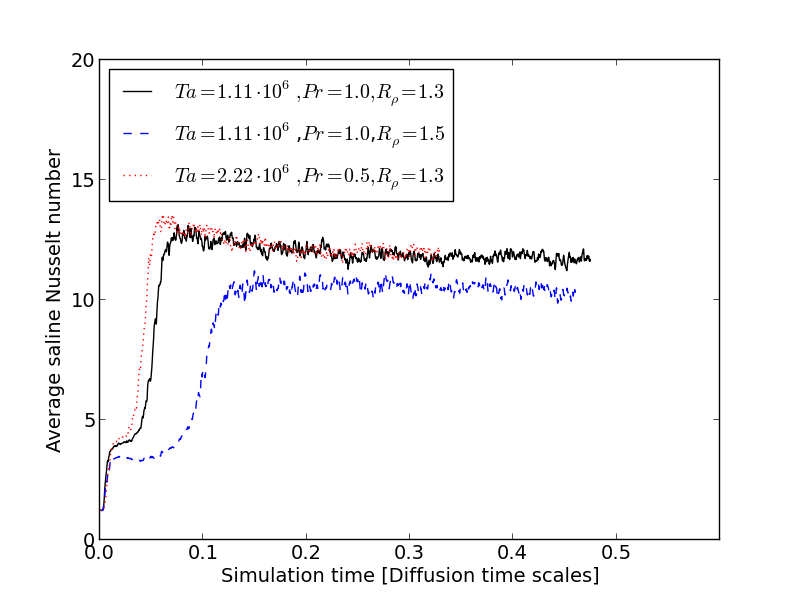}
	\end{minipage}%

	\centering
	\begin{minipage}[t]{.49\linewidth}
		\centering
		\includegraphics[width=1.0\textwidth]{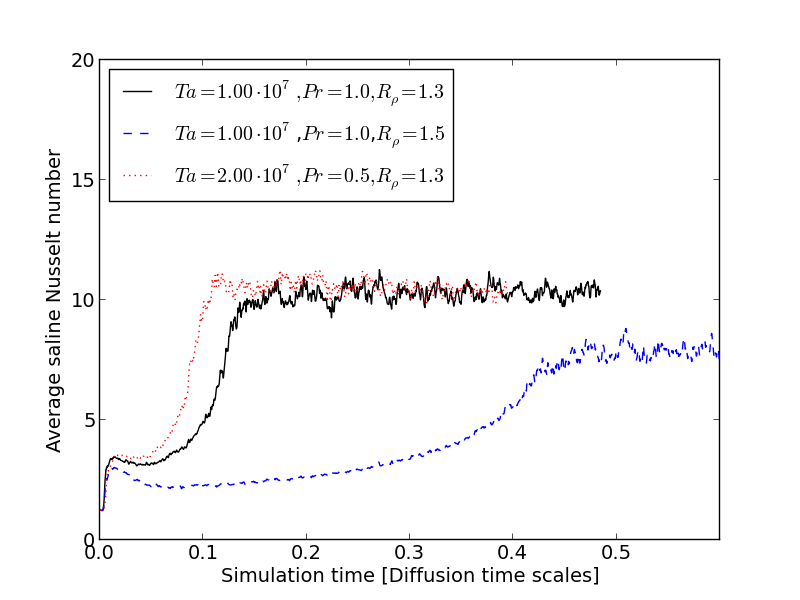}
	\end{minipage}%
	\begin{minipage}[t]{.49\linewidth}
		\centering
		\includegraphics[width=1.0\textwidth]{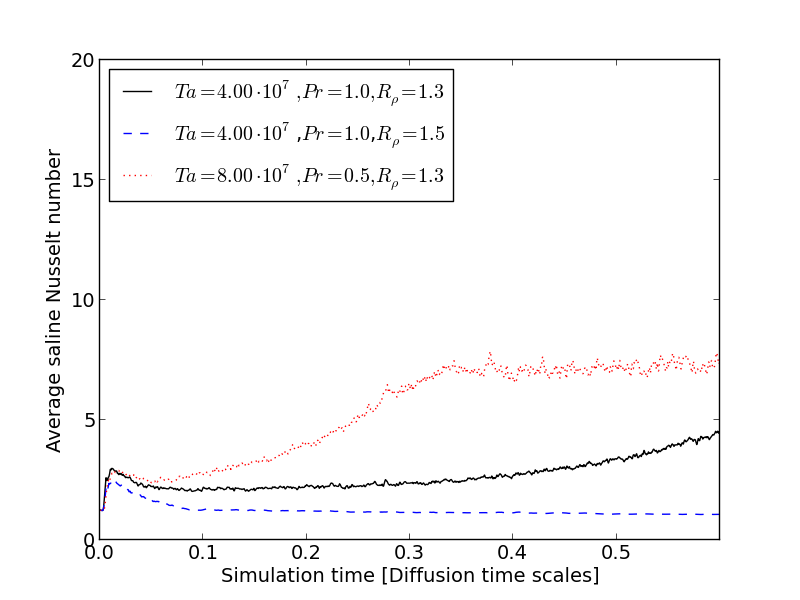}
	\end{minipage}%

	\centering
	\begin{minipage}[t]{.49\linewidth}
		\centering
		\includegraphics[width=1.0\textwidth]{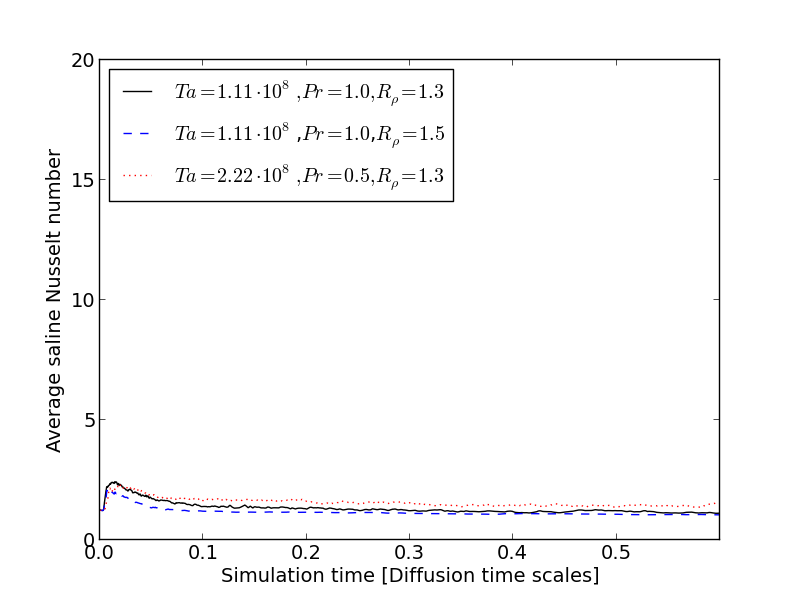}
	\end{minipage}%
	\begin{minipage}[t]{.49\linewidth}
		\centering
		\includegraphics[width=1.0\textwidth]{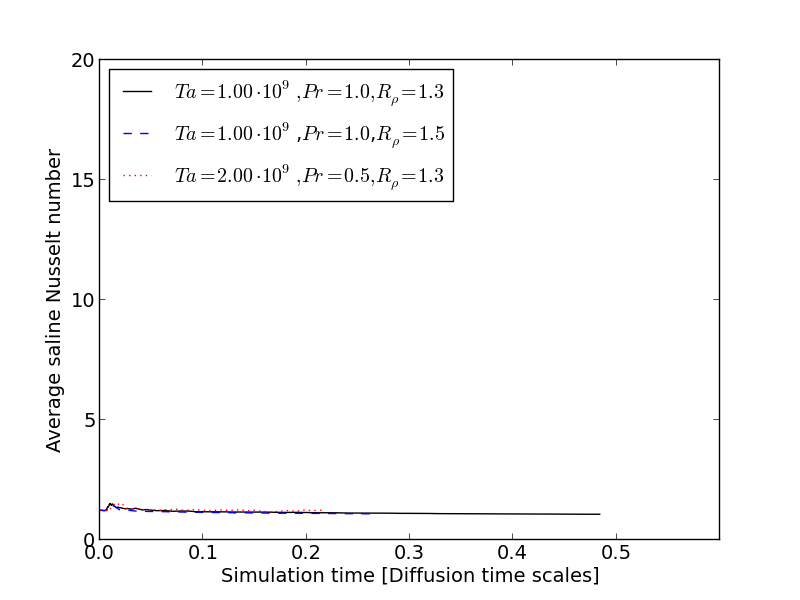}
	\end{minipage}%
	\caption{Comparison of average saline Nusselt numbers vs. simulation time at low (first row), medium (second row) and high (third row) rotation rates for a variation of Prandtl number $\Pran$ and stability factor $R_{\rho}$. }
	\label{fig:nusselt}
\end{figure}

\begin{figure}
	\centering
	\begin{minipage}[t]{.49\linewidth}
		\centering
		\includegraphics[width=1.0\textwidth]{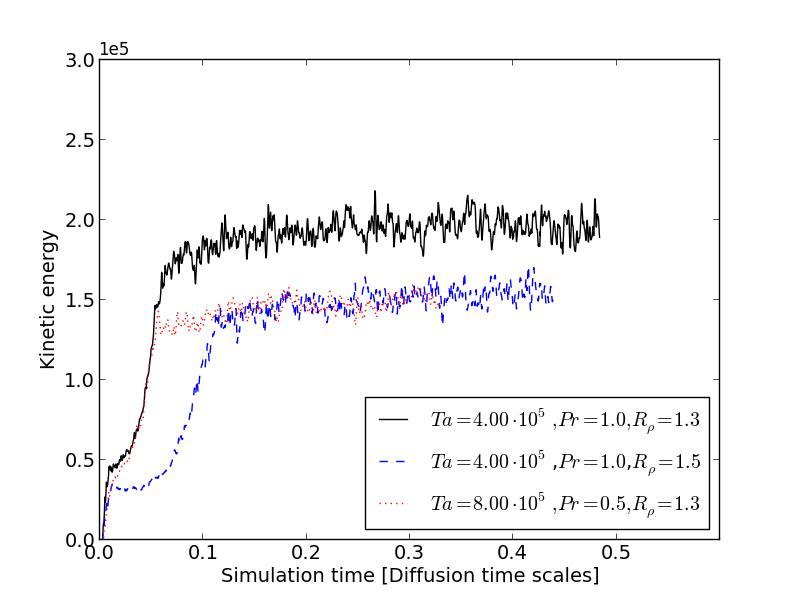}
	\end{minipage}%
	\begin{minipage}[t]{.49\linewidth}
		\centering
		\includegraphics[width=1.0\textwidth]{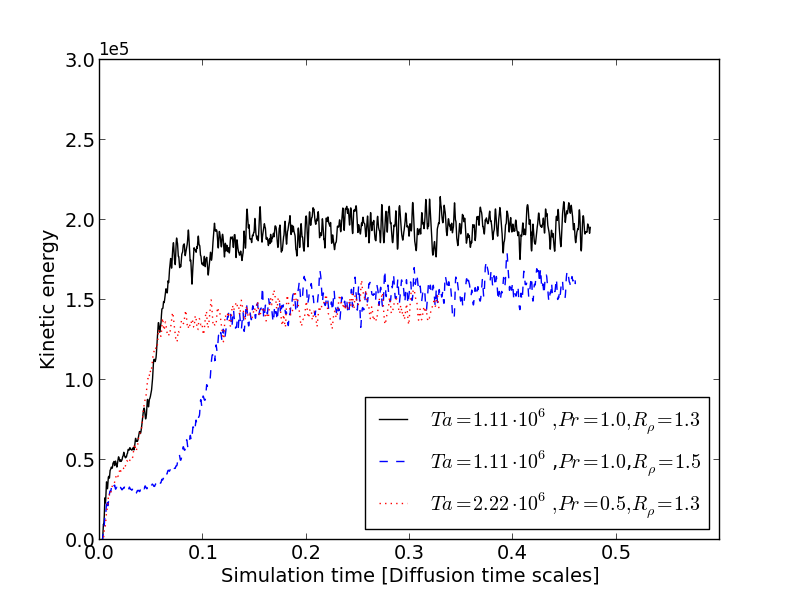}
	\end{minipage}%

	\centering
	\begin{minipage}[t]{.49\linewidth}
		\centering
		\includegraphics[width=1.0\textwidth]{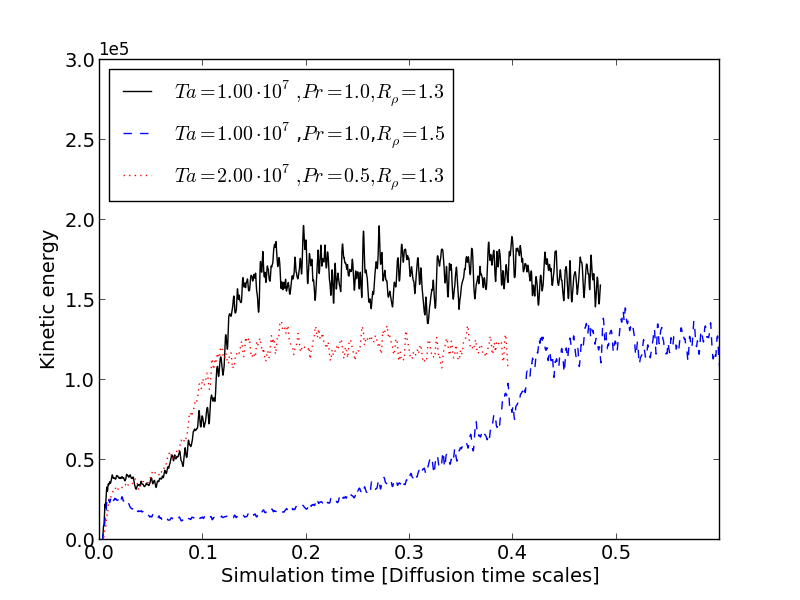}
	\end{minipage}%
	\begin{minipage}[t]{.49\linewidth}
		\centering
		\includegraphics[width=1.0\textwidth]{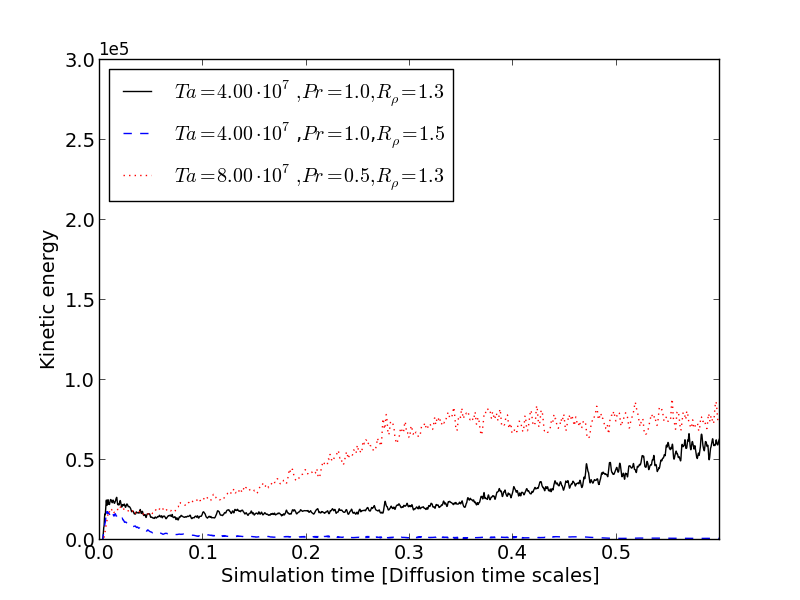}
	\end{minipage}%

	\centering
	\begin{minipage}[t]{.49\linewidth}
		\centering
		\includegraphics[width=1.0\textwidth]{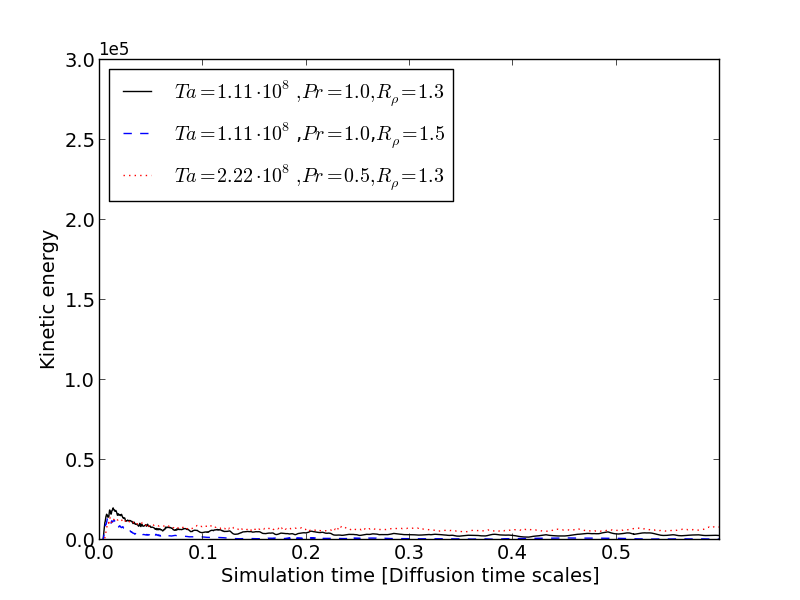}
	\end{minipage}%
	\begin{minipage}[t]{.49\linewidth}
		\centering
		\includegraphics[width=1.0\textwidth]{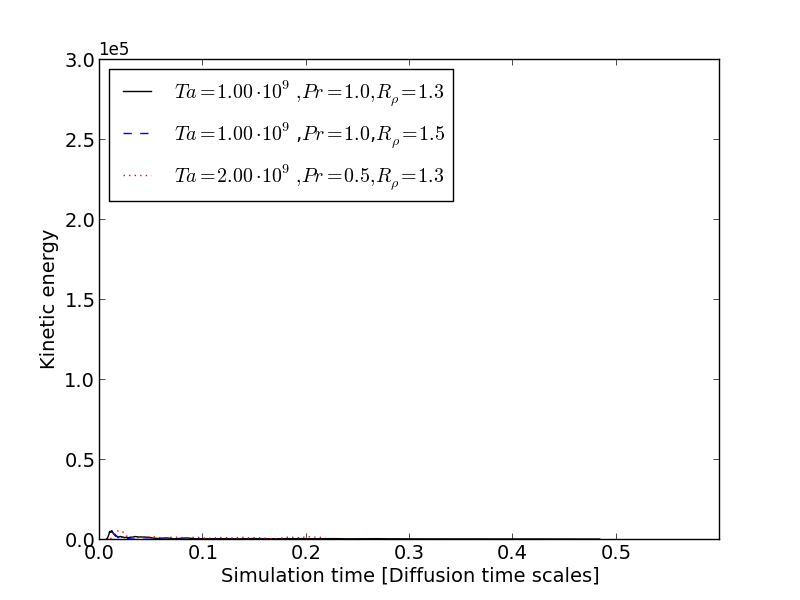}
	\end{minipage}%
	\caption{Comparison of kinetic energy vs. simulation time at low (first row), medium (second row) and high (third row) rotation rates for a variation of Prandtl number $\Pran$ and stability factor $R_{\rho}$. }
	\label{fig:kinetic}
\end{figure}

\subsubsection{Reduction of $\Pran$}
Looking at figures \ref{fig:nusselt} (a) and (b), we can observe that a reduction of $\Pran$ has no significant effect on the Nusselt number. Although the less viscous simulation reaches the asymptotic state of $\Nu\approx 6$
a bit earlier, this could very well be a coincidence. The reason for this may be that with lower viscosity there is less damping at the onset of convection so the final convective state can be reached sooner
but this is not clearly visible at low rotation rates.  \\
Comparing the kinetic energies in figures \ref{fig:kinetic} (a) and (b), a significant influence of a lower $\Pran$ is obvious: although the asymptotic limit is again reached at the same time, the value of the kinetic 
energy differs significantly. This can be explained by noting that for a more viscous system, more energy has to be stored in the convective motion in order to maintain a specific level of convection because more energy is 
converted to heat due to higher friction.

\subsubsection{Increase of $R_{\rho}$}
Even at low rotation rates, the stability parameter $R_{\rho}$ exerts a significant influence on the convective state of the system as well as on the stored kinetic energy. Figures \ref{fig:nusselt} (a) and (b) show that the system 
with a higher $R_\rho$ not only reaches a lower asymptotic limit of $\Nu\approx 5.5$ than the reference run ($\Nu \approx 6$), but it also reaches this limit at a later time which hints at a longer occurrence of a layered state.
The same holds true for the time development of kinetic energy, as is shown in figures \ref{fig:kinetic} (a) and (b).

\subsection{The effect of different $\Pran$ and $R_{\rho}$ at medium rotation rates}
\subsubsection{Reduction of $\Pran$}
Looking at figure \ref{fig:nusselt}(c) we see that a reduction of $\Pran$ to $0.5$ has a slight effect on the time when the system reaches the asymptotic limit of the average saline
Nusselt number for $\Ta=1\cdot10^7$. It has no significant effect
on the height of the limit. The effect on the kinetic energy at $\Ta = 1 \cdot 10^7$ is similar to the low rotation rate cases: 
for lower Prandtl numbers the asymptotic limit is reached a little sooner and it is lower than in the reference run.\\
At $\Ta=4\cdot10^7$ the effect of a reduced Prandtl number is clearly visible (figure \ref{fig:nusselt} d): the asymptotic limit is approached much earlier than it is in the reference run. 
We observe a similar behaviour for the kinetic energy as we did for the Nusselt numbers: the asymptotic limit for the less viscous case is reached sooner and it is lower than in the reference run.

\subsubsection{Increase of $R_{\rho}$}
Increasing $R_\rho$ has a significant effect on the convection at medium rotation rates: at $\Ta=1\cdot 10^7$ the asymptotic limit of $\Nu \approx 4$ is reached much later for $R_\rho=1.5$ than for $R_\rho=1.3$ 
(see figure \ref{fig:nusselt} c). It is also lower than the reference run ($\Nu \approx 5$). \\
A very clear effect can be seen for $\Ta=4 \cdot 10^7$ (see figure \ref{fig:nusselt} d): while for $R_\rho=1.3$ the Nusselt number and hence the heat transport by convection slowly increases, for $R_\rho=1.5$ it decreases instead 
and no convection is taking place. For kinetic energy the effect is similar (see figure \ref{fig:kinetic} d).

\subsection{The effect of different $\Pran$ and $R_{\rho}$ at high rotation rates}\label{sec:modifying_pr_and_rrho_at_higher_rotation_rates}
At high rotation rates of $\Ta=O(10^8)$ and $\Ta=O(10^9)$ convection is hindered so drastically by rotation that the effects of reducing $\Pran$ and increasing $R_\rho$ are negligible.
No convection and hence no double-diffusive convection takes place, the Nusselt number approaches $1$ (see figure \ref{fig:nusselt} f)
and the kinetic energy contained is very low.

\section{Discussion}\label{sec:discussion}
\subsection{Rotational Constraints}
\subsubsection{Characterisation through $\Ra$ and $\Ta$}
As we have seen, it also depends on the Taylor number whether heat and salt are transported by convection or diffusion only. \citet{king2013} have recently suggested that a power law constructed from the product
of Ekman and Rayleigh number can be used to describe how strong rotation affects convection. Although they have studied Rayleigh-B\'enard convection we think it is interesting to compare their results with ours.
They have found three important convection regimes: rotationally constrained convection occurs for $\Ra E^{3/2} \lesssim 10$, weakly rotating convection for $10 \lesssim \Ra E^{3/2} < \infty$
and non-rotating convection for $E^{-1} = 0$. $E$ is the Ekman number $E = \nu / (2 \Omega L^2)$, which is closely related to the Taylor number we used: $E = 1/ \sqrt{\Ta}$. Transferred to our
studies, the limiting value is given by
\begin{equation}
	\frac{\Ra}{\Ta^{3/4}}
	\label{eq:lim_value}
\end{equation}
and the three regimes are non-rotating convection for $\Ta = 0$, weakly rotating convection for $10 \lesssim \Ra / \Ta^{3/4} < \infty $ and rotationally constrained convection for
$\Ra / \Ta^{3/4} \lesssim 10 $. Which regimes our Taylor numbers belong to is seen in table \ref{tab:rot_constraint}; and indeed, the three regimes coincide very nicely with our results: 
convection is effectively prevented for $\Ta = 1.11 \cdot 10^8$ and
$\Ta = 1 \cdot 10^9$ which correspond to the regime of rotationally constrained convection of $\Ra / \Ta^{3/4} \lesssim 10 $.

However, the proposed regimes have to be expanded for the case of semiconvection. We have added an overview of the effect of a change of the density ratio $R_\rho$ on convection to table \ref{tab:rot_constraint}. We 
see that for $\Ta = 4 \cdot 10^7$ it crucially depends on $R_\rho$ if a global convective layer forms or if diffusion is the only transport mechanism of heat and salinity.
Although the system should actually be only weakly influenced by rotation since
$\Ra / \Ta^{3/4} = 19.9 > 10$, convection is completely subdued by rotation when $R_\rho = 1.5$. This suggests that the stability ratio $R_\rho$ has to enter (\ref{eq:lim_value}) in a way that a higher $R_\rho$ leads to a
reduced value (because rotation has a stronger influence):
\begin{equation}
	\frac{\Ra}{\Ta^{3/4}} \cdot f( {R_{\rho}})
	\label{eq:rrho_lim_value}
\end{equation}

This is an interesting result and worth to be studied in greater detail. In this paper, however, we restrict ourselves to the short remark that the convective regimes that \citet{king2013} proposed could also be
applicable to the case of semiconvection in a spherical shell if extended to be also a function of $R_\rho$.
\begin{table}
	\begin{center}
		\begin{tabular}{ccccc}
			 & & \multicolumn{2}{c}{Convection for} \\
			 $\Ta $ \quad   &$\Ro_{\pi / 6}$ &  $\Ra / \Ta^{3/4}$ & $R_\rho = 1.3$    &  $R_\rho = 1.5$        \\[3pt]\hline 
			$0           $ & $\infty$ &  $\infty$  & y & y \\
			$1 \cdot 10^5$ & 20 &  1780  & y & y \\
			$4 \cdot 10^5$ & 10  & 629 & y & y \\
			$1 \cdot 10^6$ & 6.0  & 292 & y & y \\ \hline
			$1 \cdot 10^7$ & 2.0  & 56.2& y & y \\
			$4 \cdot 10^7$ & 1.0  & 19.9 & y & n \\ \hline
			$1 \cdot 10^8$ & 0.6  & 9.2  & n & n \\
			$1 \cdot 10^9$ & 0.2  & 1.78& n & n 
		\end{tabular}
	\end{center}
	\caption{ Taylor numbers, Rossby numbers at colatitude $\Lambda=\pi /6$ and corresponding values of the convective regime following \cite{king2013}. The two right columns indicate if convection occurred in our simulations 
	with the indicated value for the density ratio $R_\rho$. The horizontal lines divide the table into rotationally non constrained (upper part), weakly constrained (middle part) and strongly constrained (lower part) regimes
	for $R_\rho=1.3$.
	For all simulations: $\Pran=1,\Le=0.1,\Ra = 10^7$. }
\label{tab:rot_constraint}
\end{table}
\subsubsection{Characterisation through $\Ro$}
As we have seen in the course of this paper, for $\Pran=1,\Le=0.1,R_\rho=1.3$ we have strongly constrained
convection for Taylor numbers $\Ta \in \{10^8,10^9 \}$, weakly constrained convection for $\Ta \in \{ 10^7, 4 \cdot 10^7\}$ and non constrained convection for $\Ta \in \{0, 10^5, 4 \cdot 10^5, 10^6 \}$. This is indicated by the
horizontal lines in table \ref{tab:rot_constraint}.
It is interesting to compare the Rossby numbers from table \ref{tab:taylor_rossby} to the constraint that rotation exerts on the flow in our simulations. For the stability ratio $R_{\rho} = 1.3$, we get the result that if 
$\Ro<0.5$ in the bulk of the shell (indicated by the Rossby number at colatitude $\Lambda=\pi /6$ in table \ref{tab:rot_constraint}), 
we have the case of rotationally strongly constrained convection. The weakly constrained (or transition) cases correspond to $0.5 < \Ro < 2$ while the non constrained cases correspond to
$\Ro > 2$. 

It is interesting to note that the Reynold stress correlations investigated in \citet{chan2001} and the structure of the temperature field shown in \citet{chan2007}, both times for so-called f-box simulations of rotating
convection with uniform composition and a fully compressible flow, 
show a similar transition region at Coriolis numbers (which are defined as $1/\Ro$) 
that correspond to exactly the same regime of Rossby numbers as in our case with $R_{\rho}=1.3$: a transition region for $0.5< \Ro<1$ and a rotation dominated
flow for $\Ro < 0.5$, at equatorial co-latitude respectively.

However, as in the case of King et al.'s model, the Rossby number alone seems to be insufficient for figuring out the effect of rotation on semiconvection. Again, $R_\rho$ presents itself to be a crucial factor.
For $R_\rho = 1.5$ we have a higher influence of rotation than for $R_\rho=1.3$. An increase to $R_\rho=1.5$ seems to shift the Rossby numbers by a factor of about $0.5$, meaning that the effective Rossby number for $\Ta=4 \cdot
10^7$ would be $\Roeff=0.5$. With this value, it enters the rotationally strongly constrained regime. And indeed, for $\Ta=4 \cdot 10^7$ and $R_{\rho}=1.5$ we have no convection (see figure \ref{fig:nusselt}). So the stability ratio
also affects the Rossby number in a way that a higher $R_\rho$ leads to a lower effective Rossby number \Roeff:
\begin{equation}
 \Roeff = \frac{\Ro}{f(R_\rho)}.
\end{equation}

\subsection{The relationship between $\Nus$ and $\Nut$}
\subsubsection{The relationship between $\Nus$ and $\Nut$ for $\Ta=0$}

There exist different models for the relationship between the thermal and saline Nusselt number.
According to \citet{spruit_theory_2013} they are related via
\begin{equation}
	\Nus - 1 = \frac{q}{\Le^{1/2} R_\rho} (\Nut - 1) 
	\label{eq:spruit_nusselt_theory}
\end{equation}
for $R_\rho < \Le^{-1/2}$. $q$ is a fit parameter. According to (32) of \citet{Rosenblum2011} the relationship is
\begin{equation}
	\Nus - 1 \approx \frac{1}{\Le R_\rho } (\Nut - 1),
	\label{eq:rosenblums_theory}
\end{equation}
which is especially for very low Lewis numbers in strong contrast to theoretical results from the linear stability theory.
According to (42) and (43) of \citet{wood_2013}, the relation is given by 
\begin{equation}
	\Nus - 1 = \frac{B}{A}  \frac{\Pran^{1/12}}{\Le} \Rat^{0.37-1/3} (\Nut - 1).
	\label{eq:woods_theory}
\end{equation}
The latter is given for $\Pran \ll 1$ which is not the case in our simulations, so a deviation can be expected. Typical values for $A$ and $B$ are given as $A\approx 0.1$ and $B \approx 0.03$, so 
$B/A \approx 0.3$.

\begin{figure}
	\begin{center}
	\begin{minipage}[]{0.7\linewidth}
		\includegraphics[width=\textwidth]{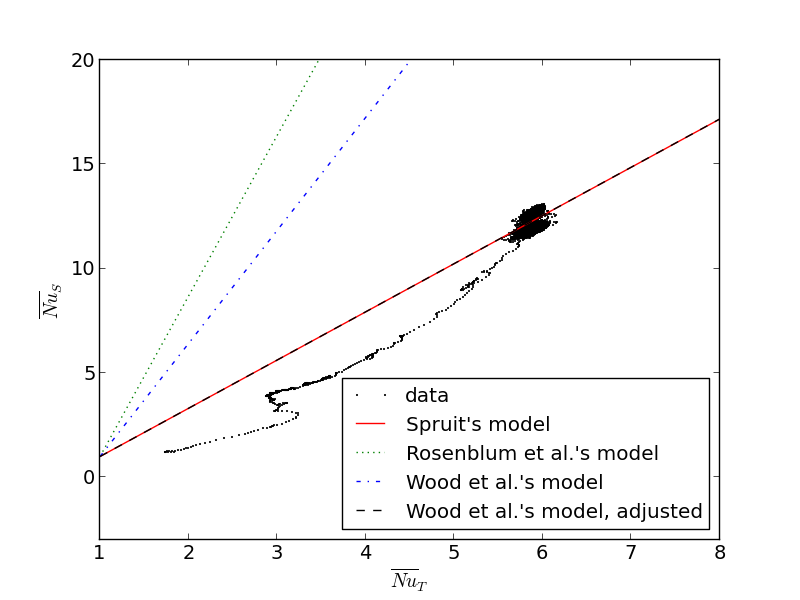}
	\end{minipage}
	\caption{Average saline Nusselt number vs. average thermal Nusselt number for $\Ta=0$. The black dots are data points from our simulation, the (red) solid line is a plot of
	(\ref{eq:spruit_nusselt_theory}) with $q=0.95$, 
	the (green) dotted line is a plot of (\ref{eq:rosenblums_theory}), the (blue) dash-dotted line is a plot of (\ref{eq:woods_theory}) with $B/A=0.3$, the (black)
	dashed line a plot of (\ref{eq:woods_theory}) with $B/A=0.128$.}
	\label{fig:theory_data}
	\end{center}
\end{figure}

These three models are tested here against our results of average thermal and saline Nusselt numbers.
The result for the simulation without rotation is seen in figure \ref{fig:theory_data}.
We see that Spruit's theoretical prediction (solid line in figure \ref{fig:theory_data}) 
lies in the vicinity of the data points but does not exactly reproduce them except in one area where there is a big amount of data points at $\Nut \approx 6$. 
This area of abundant data represents the system, when it has reached the statistically stable 
end state. The line of data points, on the other hand, represents the system while it is relaxing to its end state. 
Rosenblum et al.'s model (dotted line in figure \ref{fig:theory_data}) does not fit our data. Since it does not have a fitting parameter, it cannot be adjusted to fit the data, either.
Wood et al.'s model does not fit our data with their proposed values for $A$ and $B$. It does, however, fit the data equally good as Spruit's model does when adjusting $A$ and $B$ accordingly.
We can therefore conclude that Spruit's model and the adjusted version of Wood et al.'s model both make successful predictions for the ratio of saline and thermal Nusselt numbers in the parameter range that we simulated in the
non-rotating case.
Albeit, they do so only after the system has reached its equilibrium state. 

Taking Spruit's model as a starting point, we calculate the interval that $\Nus/\Nut$ has to lie in.
Starting from (\ref{eq:spruit_nusselt_theory}) after a few elementary transformations we get
\begin{equation}
	\frac{\Nus}{\Nut} = \frac{q}{\Le^{1/2} R_\rho} - \frac{1}{\Nut} \left( \frac{q}{Le^{1/2} R_\rho} - 1 \right) \equiv g(\Nut)  
\end{equation}
The smallest possible value of $\Nut$ is one, which corresponds to the case of pure diffusion: 
\begin{equation}
	g(1) = \frac{q}{\Le^{1/2} R_\rho} - \frac{q}{Le^{1/2} R_\rho} + 1  = 1.
\end{equation}
In the limit of $\Nut \rightarrow \infty$ we get
\begin{equation}
	\mathrm{lim}_{\Nut \rightarrow \infty} g(\Nut) = \frac{q}{Le^{1/2} R_\rho}.
\end{equation}
Provided that $q \ge \Le^{1/2}R_\rho$ the ratio of Nusselt numbers $g(\Nut)$ is therefore bounded by
\begin{equation}
	1 \le \frac{\Nus}{\Nut} \le \frac{q}{Le^{1/2} R_\rho}.
\end{equation}
In our reference case, where $Le=0.1$ and $R_\rho=1.3$ this gives
\begin{equation}
	1 \le \frac{\Nus}{\Nut} \le 2.43 \, q.
\end{equation}
Setting the fit parameter $q=0.95$, which is the value that fits the data in figure \ref{fig:theory_data}, we end up with
\begin{equation}
	1 \le \frac{\Nus}{\Nut} \le 2.31
\end{equation}

\begin{figure}
	\begin{minipage}{0.5\linewidth}
		\begin{center}
			\includegraphics[width=\textwidth]{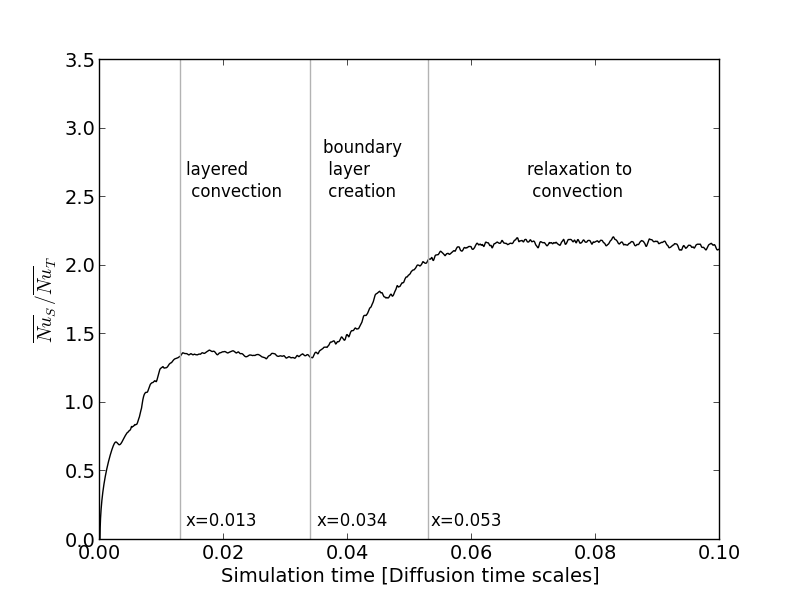}
		\end{center}
	\end{minipage}
	\begin{minipage}{0.5\linewidth}
		\begin{center}
			\includegraphics[width=\textwidth]{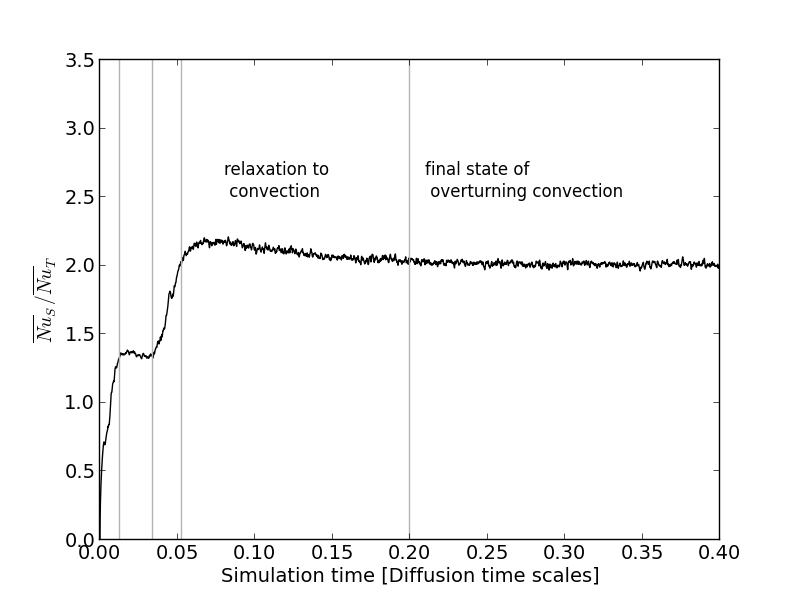}
		\end{center}
	\end{minipage}
	\caption{Ratio of average saline and thermal Nusselt number vs. simulation time for up to $t=0.1$ (left) and $t=0.4$ for $\Ta=0$. The state of the system can be 
	separated into different regions of sharply rising and slowly declining ratio of Nusselt numbers that correspond to states of layered convection, boundary layer creation and overturning convection.}
	\label{fig:ratio_of_nusselts}
\end{figure}

Our data confirms this for all times except for the initial plume phase, as can be observed from figures \ref{fig:thrice_taylor0} and \ref{fig:ratio_of_nusselts}. The maximal value of $\overline{\Nus}/\overline{\Nut}$ is $\approx 2.25$.
The plots 
show another interesting result.
It appears that the ratio of Nusselt numbers is a good classification for the state of the flow.
After the plumes have broken, a convective layer is established (at $t\approx 0.013$ in 
figure \ref{fig:ratio_of_nusselts}). The thickness of the layer $d_s$ increases with time until the top of the layer reaches the upper boundary of the shell (at $t\approx 0.034$ ). Then, thermal and saline diffusive 
boundary layers at the shell boundary are established. This is indicated by a rise of $\overline{\Nus} / \overline{\Nut}$. We suspect that if we had a second layer on top, 
these would then start to merge at this point. But since we have imposed boundary 
conditions there, the state of the system relaxes to a homogeneous convective layer embedded between diffusive transition ranges at both top and bottom, a state that is reached at $t \approx 0.2$ .

We note here that during layer formation and extension $q$ and, likewise, $A$ and $B$ in the models of Spruit and Wood et al., respectively, may not remain constant and their values may not be the same for differently sized stacks
of layers or during a ``merging process'' or for different boundary conditions.

If and which influence rotation and a change of $\Pran$ and $R_\rho$ have on $\overline{\Nus} / \overline{\Nut}$ will be investigated next. 

\subsubsection{Influence of rotation on the relationship between $\Nus$ and $\Nut$}\label{sss:ratio_nusselts_rotation}
\begin{figure}
	\begin{minipage}{0.5\linewidth}
		\begin{center}
			\includegraphics[width=\textwidth]{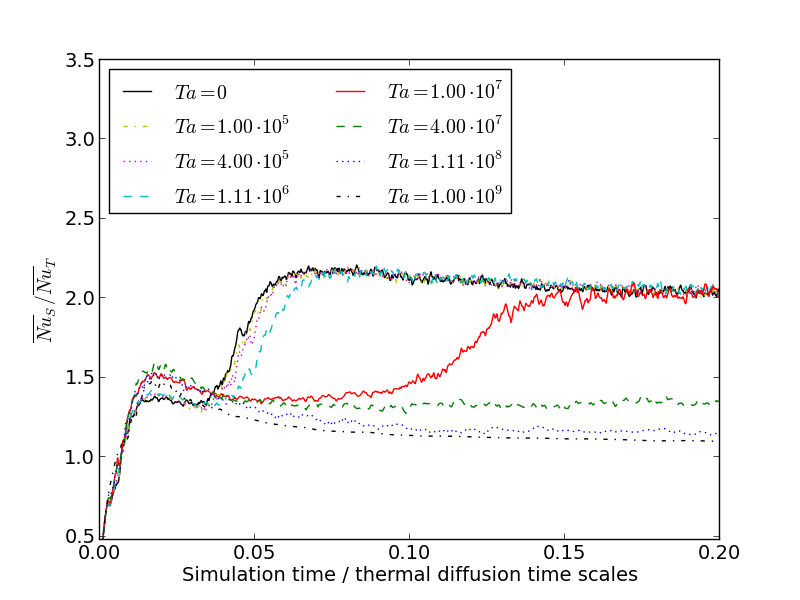}
		\end{center}
	\end{minipage}
	\begin{minipage}{0.5\linewidth}
		\begin{center}
			\includegraphics[width=\textwidth]{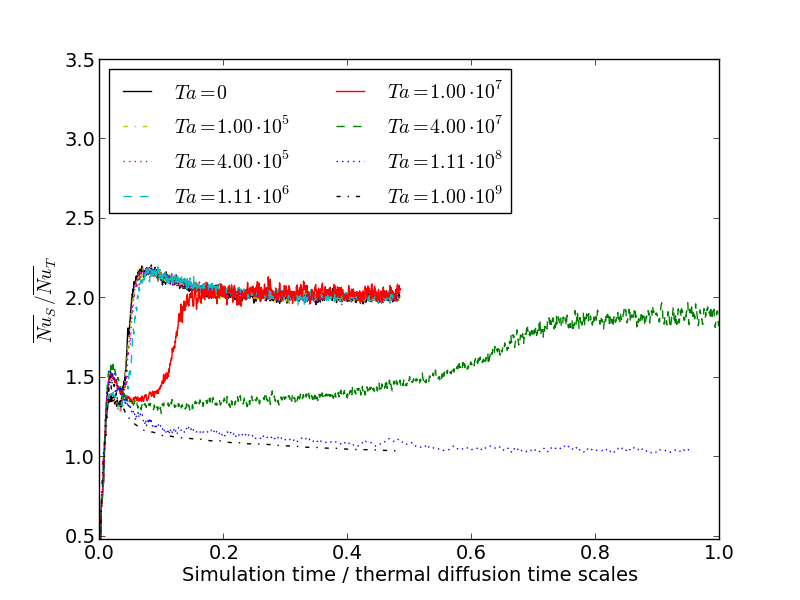}
		\end{center}
	\end{minipage}
	\begin{minipage}{0.5\linewidth}
		\begin{center}
			\includegraphics[width=\textwidth]{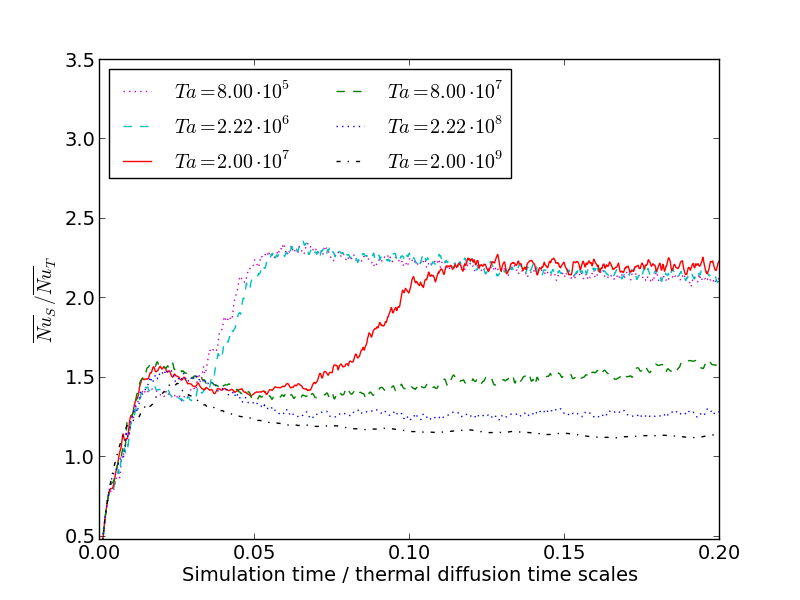}
		\end{center}
	\end{minipage}
	\begin{minipage}{0.5\linewidth}
		\begin{center}
			\includegraphics[width=\textwidth]{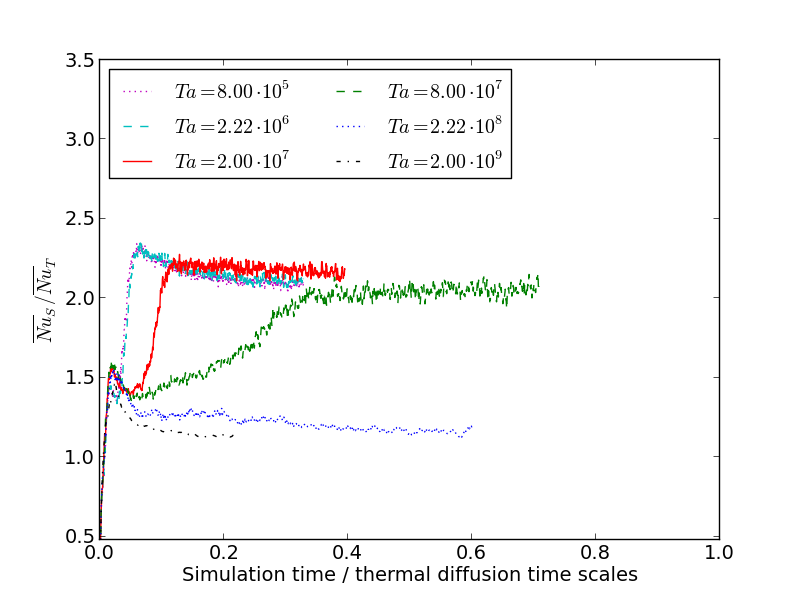}
		\end{center}
	\end{minipage}
	\begin{minipage}{0.5\linewidth}
		\begin{center}
			\includegraphics[width=\textwidth]{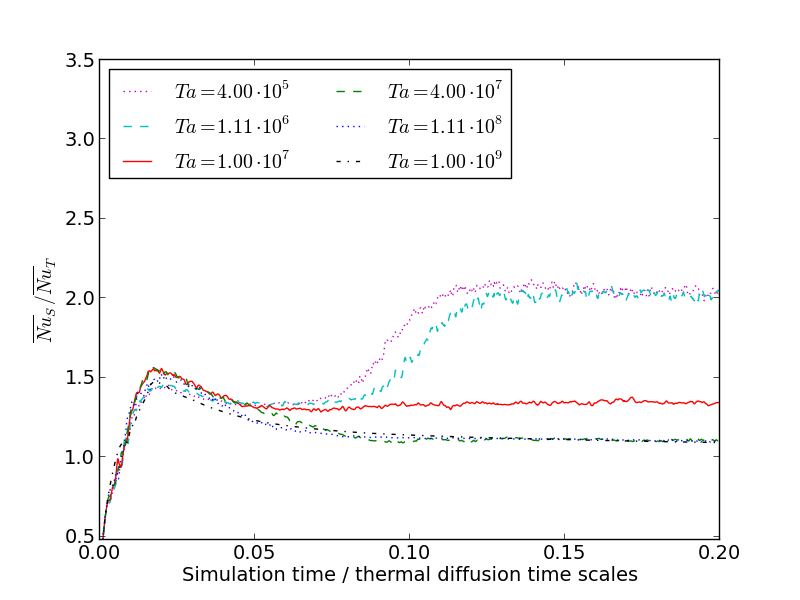}
		\end{center}
	\end{minipage}
	\begin{minipage}{0.5\linewidth}
		\begin{center}
			\includegraphics[width=\textwidth]{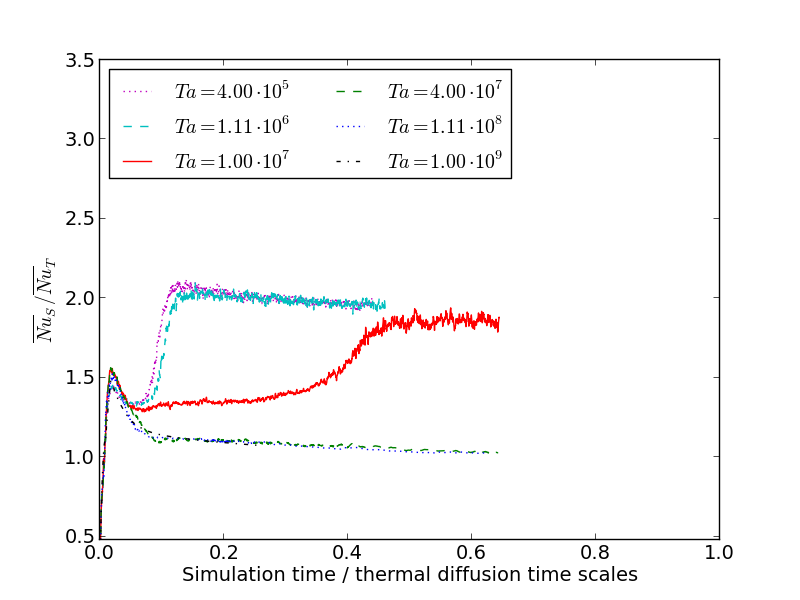}
		\end{center}
	\end{minipage}
	\caption{The ratio of saline and thermal Nusselt numbers for the three parameter pairs $(\Pran=1, R_\rho=1.3)$ (upper row), $(\Pran=0.5, R_\rho=1.3)$ (middle row) and $(\Pran=1, R_\rho=1.5)$ (lower row) for simulation
	times up to $t=0.02$ (left column) and $t=1$ (right column).}
	\label{fig:nusselt_relation}
\end{figure}

The first row of figure \ref{fig:nusselt_relation} shows $\overline{\Nus} / \overline{\Nut}$ against the simulation time for different rotation rates up to $t=0.2$ (left) and up to $t=1$ (right).
For no rotation, at low simulation times, the ratio of saline and thermal Nusselt numbers rises monotonously with a high slope until it relaxes to a slightly falling ``plateau of (slowly growing) layered convection''
only to sharply rise again and relax to a statistically continuous value. Rotation seems to stretch the plateau which exists while the thickness $d_s$ of the layer grows and turns it into 
more of a sink and makes it much wider for higher rotation rates. For the highest rotation rates,
the ones where the final state is one of diffusion, there is no plateau/sink at all, the ratio of Nusselt numbers falls monotonously and approaches one. 

The maximum value for $\overline{\Nus} / \overline{\Nut}$ does not change for $\Ta \le 1.11 \cdot 10^6$. For $\Ta=10^7$ it decreases from $\approx 2.25$ obtained for all lower Taylor numbers to $\approx 2$. For
$Ta=4 \cdot 10^7$ it decreases further to $\approx 1.8$. For $\Ta=1.11 \cdot 10^8$ the ratio of Nusselt numbers does not reach its global maximum after $t\approx 0.05$ but before. This is before the system is in statistic
equilibrium. When it reaches the equilibrium state, convection is suppressed and both Nusselt numbers tend to one. The same is true for $\Ta=10^9$. \\

Since fast rotation does have a significant effect on the ratio of Nusselt numbers and hence the flux ratio, a measure of the rate of rotation has to enter the theoretical prediction (\ref{eq:spruit_nusselt_theory}) or the fit
formula (\ref{eq:woods_theory}).

\subsubsection{Modifying $\Pran$ and $R_\rho$}
Looking at the middle and last row of figure \ref{fig:nusselt_relation} we get the same result that was seen in chapters \ref{sec:modifying_pr_and_rrho} to \ref{sec:modifying_pr_and_rrho_at_higher_rotation_rates}. Lowering the Prandtl 
number decreases the influence 
that a higher rotation rate has on the stability of
semiconvection. While for $\Ta=4\cdot 10^7$ the asymptotic state of the ratio of Nusselt numbers occurs at $t \approx 0.7$, for a Prandtl number half as high, the corresponding case with $\Ta=8\cdot 10^7$ reaches the asymptotic state 
already at $t \approx 0.35$. So the phases of layered convection and the creation of a diffusive boundary layer happen on a shorter time scale for lower Prandtl numbers. Likewise, increasing $R_{\rho}$ to $1.5$ on the other hand
slows down the time development for $\Ta = 10^7$ and turns $\Ta = 10^7$ into a diffusive case where $\Nus/\Nut$ drops to 1 after an early maximum value. The saturation of $\Nus/\Nut$ for the non-diffusive cases occurs at the same
time as for $\Nus$.

\subsection{The influence of rotation on the lifetime of a layer}
Next, we take note of the time at which the 
thermal Nusselt number reaches the asymptotic value. We chose the thermal Nusselt number because it has the sharpest kink when reaching the statistically stable state. 
The results are summarised in table \ref{tab:asymptotic_times}. The times were taken from figure \ref{fig:dts01udts1}.
\begin{table}
	\begin{center}
		\begin{tabular}{rccc}
$\Ta \cdot \Pran$  \quad  & \multicolumn{3}{c}{$t_{\mathrm{asymptotic}}$}  \\[3pt]\hline 
 & \multicolumn{1}{l}{ $\quad \Pran = 1, R_\rho = 1.3  \quad $} & $\Pran = 1, R_\rho = 1.5$ \quad & $\Pran = 0.5, R_\rho = 1.3$ \quad \\ \hline
\multicolumn{1}{r}{0} & 0.056 & n/a & n/a \\
$1 \cdot 10^5$ & 0.056 & n/a & n/a \\
$4 \cdot 10^5$ & 0.059 & \multicolumn{1}{c}{0.110} & \multicolumn{1}{c}{0.046} \\
$1.11 \cdot 10^6$ & 0.060 & \multicolumn{1}{c}{0.124} & \multicolumn{1}{c}{0.048} \\
$1 \cdot 10^7$ & 0.132 & \multicolumn{1}{c}{0.428} & \multicolumn{1}{c}{0.106} \\
$4 \cdot 10^7$ & 0.724 & $\infty$ & \multicolumn{1}{c}{0.328} \\
$1.11 \cdot 10^8$ & \multicolumn{1}{c}{$\infty$} & $\infty$ & $\infty$ \\
$1 \cdot 10^9$ & \multicolumn{1}{c}{$\infty$} & $\infty$ & $\infty$ \\
\end{tabular}
	\end{center}
\caption{Time in thermal diffusion time scales when thermal Nusselt number reaches the statistically stable asymptotic state. n/a means that we did not run simulations for these parameters. $\infty$ means that there is no 
statistically stable convective state for these parameters.}
\label{tab:asymptotic_times}
\end{table}
Obviously, increasing $R_{\rho}$ delays the time development of $\Nut$ while lowering $\Pran$ accelerates it.
But since these are far too few data points to make a sound assumption about an underlying law we restrict ourselves to just listing them.

\section{Summary and Outlook}\label{sec:conclusion}
We have studied the influence of rotation on semiconvection in a three dimensional spherical shell. First, we have run simulations without rotation with $\Pran=1$ and a stability ratio $R_\rho=1.3$ 
to set up a reference calculation.
Then, we compared simulations with different rates of rotation to the non-rotating case and compared the values for thermal and saline Nusselt numbers and kinetic energies. 
We concluded that slow rotation has hardly any influence on layer formation while fast rotation suppresses convective transport completely. At intermediate rotation rates
the temporal development of layers may take an entire thermal diffusion time scale at some critical \Ta. This is why short simulation times can be highly misleading as there may be a long relaxation phase. 
At low rotation rates 
the relaxation of $\Nut$ to its quasi-equilibrium value occurs on a much smaller timescale that for $\Nus$.
Furthermore, for higher $\Ta$ the equilibrium state is no longer quasi-adiabatic but becomes increasingly more 
superadiabatic with ever more extended, diffusive transition regions at the layer boundaries.

We have also compared results from simulations with a modified Prandtl number or a modified stability ratio $R_\rho$ and conclude that
the critical value of the Taylor number $\Tac$ at which convection is suppressed
depends weakly on $\Pran$ and strongly on $R_{\rho}$. For lower $\Pran$ the equilibrium state requires less (turbulent) kinetic energy, since less viscous friction occurs which converts kinetic energy into heat. 
For higher $R_{\rho}$ the maximum $\Ta$ for which convection
develops, drops, so the critical $R_{\rho}$ in the sense of \citet{spruit_theory_2013} and \citet{radko2003} 
should depend on $\Ta$. We have also compared our results with the three regimes of rotational constraints that \citet{king2013} suggested and conclude that they are a valid means of classifying the effect of rotation on 
semiconvection but have to be extended by a dependence on $R_\rho$. Similarly, if using the Rossby number as a means for the influence of rotation it has to be extended by a dependence on $R_\rho$ as well.

We have studied the relationship of $\Nut$ and $\Nus$ which seems to be a good indication for the state of the flow in terms of layered convection, boundary layer creation, relaxation to thoroughly mixed convection and the final
state of overturning convection.
For the case without rotation, we compared our data with model predictions by \citet{spruit_theory_2013}, \citet{Rosenblum2011} and \citet{wood_2013}.
For the equilibrium state of the system, Spruit's model fits our data perfectly, while Wood et al.'s model does so only after readjusting their fitting parameters. Rosenblum et al.'s model does not deliver a satisfactory fit to our
data. For the rotating case, the models have to be readjusted, however, since fast rotation is found to significantly affect the ratio of Nusselt numbers.

Our work is a step on the ladder of correctly understanding the influence of double-diffusive convection on the heat and solute transport in rapidly rotating systems like some stars and planets. We already 
know that semiconvection significantly constrains heat transport. If we now respect that high rates of rotation can further decrease the effective heat and solute fluxes, the assumed 
overall heat flux in a rapidly rotating system undergoing double-diffusive convection has to be lowered even further when modeling planets or stars. Future works can include a more precise study of the influence 
of rotation on the lifetime of a layer and an investigation of the opposite regime of double-diffusive convection: that of salt-fingering. \\

PB \&  FK are grateful to financial support from the Austrian Science Fund (FWF) through project P25229-N27. RH is supported by STFC grant ST/K000853/1.

\bibliographystyle{chicago}

\begin{thebibliography}{}

\bibitem[\protect\citeauthoryear{{Castaing}, {Gunaratne}, {Kadanoff},
  {Libchaber}, and {Heslot}}{{Castaing} et~al.}{1989}]{Castaing1989}
{Castaing}, B., G.~{Gunaratne}, L.~{Kadanoff}, A.~{Libchaber}, and F.~{Heslot}
  (1989).
\newblock {Scaling of hard thermal turbulence in Rayleigh-Benard convection}.
\newblock {\em Journal of Fluid Mechanics\/}~{\em 204}, 1--30.

\bibitem[\protect\citeauthoryear{{Chabrier} and {Baraffe}}{{Chabrier} and
  {Baraffe}}{2007}]{chabrier2007}
{Chabrier}, G. and I.~{Baraffe} (2007).
\newblock {Heat Transport in Giant (Exo)planets: A New Perspective}.
\newblock {\em The Astrophysical Journal Letters\/}~{\em 661}, L81--L84.

\bibitem[\protect\citeauthoryear{{Chan}}{{Chan}}{2001}]{chan2001}
{Chan}, K.~L. (2001).
\newblock {Rotating Convection in F-Planes: Mean Flow and Reynolds Stress}.
\newblock {\em The Astrophysical Journal\/}~{\em 548}, 1102--1117.

\bibitem[\protect\citeauthoryear{{Chan}}{{Chan}}{2007}]{chan2007}
{Chan}, K.~L. (2007).
\newblock {Rotating convection in f-boxes: Faster rotation}.
\newblock {\em Astronomische Nachrichten\/}~{\em 328}, 1059.

\bibitem[\protect\citeauthoryear{Flanagan, Lefler, and Radko}{Flanagan
  et~al.}{2013}]{Flanagan20132466}
Flanagan, J., A.~Lefler, and T.~Radko (2013).
\newblock Heat transport through diffusive interfaces.
\newblock {\em Geophysical Research Letters\/}~{\em 40\/}(10), 2466--2470.

\bibitem[\protect\citeauthoryear{Hollerbach}{Hollerbach}{2000}]{hollerbach_200%
0}
Hollerbach, R. (2000).
\newblock A spectral solution of the magneto-convection equations in spherical
  geometry.
\newblock {\em International Journal for Numerical Methods in Fluids\/}~{\em
  32\/}(7), 773--797.

\bibitem[\protect\citeauthoryear{{Huppert} and {Moore}}{{Huppert} and
  {Moore}}{1976}]{huppert1976}
{Huppert}, H.~E. and D.~R. {Moore} (1976).
\newblock {Nonlinear double-diffusive convection}.
\newblock {\em Journal of Fluid Mechanics\/}~{\em 78}, 821--854.

\bibitem[\protect\citeauthoryear{{Huppert} and {Turner}}{{Huppert} and
  {Turner}}{1981}]{huppert_1981}
{Huppert}, H.~E. and J.~S. {Turner} (1981).
\newblock {Double-diffusive convection}.
\newblock {\em Journal of Fluid Mechanics\/}~{\em 106}, 299--329.

\bibitem[\protect\citeauthoryear{{Kerr}}{{Kerr}}{1996}]{Kerr1996}
{Kerr}, R.~M. (1996).
\newblock {Rayleigh number scaling in numerical convection}.
\newblock {\em Journal of Fluid Mechanics\/}~{\em 310}, 139--179.

\bibitem[\protect\citeauthoryear{{King}, {Stellmach}, and {Buffett}}{{King}
  et~al.}{2013}]{king2013}
{King}, E.~M., S.~{Stellmach}, and B.~{Buffett} (2013).
\newblock {Scaling behaviour in Rayleigh-B{\'e}nard convection with and without
  rotation}.
\newblock {\em Journal of Fluid Mechanics\/}~{\em 717}, 449--471.

\bibitem[\protect\citeauthoryear{Leconte and Chabrier}{Leconte and
  Chabrier}{2012}]{Leconte2012}
Leconte, J. and G.~Chabrier (2012).
\newblock A new vision of giant planet interiors: Impact of double diffusive
  convection.
\newblock {\em Astronomy \& Astrophysics\/}~{\em 540}, A20.

\bibitem[\protect\citeauthoryear{{Leconte} and {Chabrier}}{{Leconte} and
  {Chabrier}}{2013}]{Leconte2013}
{Leconte}, J. and G.~{Chabrier} (2013).
\newblock {Layered convection as the origin of Saturn's luminosity anomaly}.
\newblock {\em Nature Geoscience\/}~{\em 6}, 347--350.

\bibitem[\protect\citeauthoryear{{Ledoux}}{{Ledoux}}{1947}]{ledoux_1947}
{Ledoux}, P. (1947).
\newblock {Stellar Models with Convection and with Discontinuity of the Mean
  Molecular Weight}.
\newblock {\em The Astrophysical Journal\/}~{\em 105}, 305.

\bibitem[\protect\citeauthoryear{Mirouh, Garaud, Stellmach, Traxler, and
  Wood}{Mirouh et~al.}{2012}]{mirouh2012}
Mirouh, G.~M., P.~Garaud, S.~Stellmach, A.~L. Traxler, and T.~S. Wood (2012).
\newblock A new model for mixing by double-diffusive convection
  (semi-convection). i. the conditions for layer formation.
\newblock {\em The Astrophysical Journal\/}~{\em 750\/}(1), 61.

\bibitem[\protect\citeauthoryear{Net, Garcia, and S\'anchez}{Net
  et~al.}{2012}]{Net2012}
Net, M., F.~Garcia, and J.~S\'anchez (2012).
\newblock Numerical study of the onset of thermosolutal convection in rotating
  spherical shells.
\newblock {\em Physics of Fluids\/}~{\em 24\/}(6).

\bibitem[\protect\citeauthoryear{{O'}Rourke and Stevenson}{{O'}Rourke and
  Stevenson}{2014}]{ORourke2014}
{O'}Rourke, J.~G. and D.~J. Stevenson (2014).
\newblock Stability of ice/rock mixtures with application to a partially
  differentiated titan.
\newblock {\em Icarus\/}~{\em 227\/}(0), 67 -- 77.

\bibitem[\protect\citeauthoryear{Radko}{Radko}{2003}]{radko2003}
Radko, T. ({2003}).
\newblock {A mechanism for layer formation in a double-diffusive fluid}.
\newblock {\em Journal of Fluid Mechanics\/}~{\em {497}}, {365--380}.

\bibitem[\protect\citeauthoryear{{Rosenblum}, {Garaud}, {Traxler}, and
  {Stellmach}}{{Rosenblum} et~al.}{2011}]{Rosenblum2011}
{Rosenblum}, E., P.~{Garaud}, A.~{Traxler}, and S.~{Stellmach} (2011).
\newblock {Turbulent Mixing and Layer Formation in Double-diffusive Convection:
  Three-dimensional Numerical Simulations and Theory}.
\newblock {\em The Astrophysical Journal\/}~{\em 731}, 66.

\bibitem[\protect\citeauthoryear{Schmid, Busbridge, and Wüest}{Schmid
  et~al.}{2010}]{Schmid2010225}
Schmid, M., M.~c. Busbridge, and A.~b. Wüest (2010).
\newblock Double-diffusive convection in {L}ake {K}ivu.
\newblock {\em Limnology and Oceanography\/}~{\em 55\/}(1), 225--238.

\bibitem[\protect\citeauthoryear{{Schwarzschild} and
  {H{\"a}rm}}{{Schwarzschild} and {H{\"a}rm}}{1958}]{schwarzschild_haerm_1958}
{Schwarzschild}, M. and R.~{H{\"a}rm} (1958).
\newblock {Evolution of Very Massive Stars.}
\newblock {\em The Astrophysical Journal\/}~{\em 128}, 348.

\bibitem[\protect\citeauthoryear{{Spruit}}{{Spruit}}{2013}]{spruit_theory_2013}
{Spruit}, H.~C. (2013).
\newblock {Semiconvection: theory}.
\newblock {\em Astronomy \& Astrophysics\/}~{\em 552}, A76.

\bibitem[\protect\citeauthoryear{{Stevenson}}{{Stevenson}}{1982}]{stevenson198%
2a}
{Stevenson}, D.~J. (1982).
\newblock {Formation of the giant planets}.
\newblock {\em Planetary and Space Science\/}~{\em 30}, 755--764.

\bibitem[\protect\citeauthoryear{{Stommel}, {Arons}, and {Blanchard}}{{Stommel}
  et~al.}{1956}]{stommel1956}
{Stommel}, H., A.~B. {Arons}, and D.~{Blanchard} (1956).
\newblock {An oceanographical curiosity: the perpetual salt fountain}.
\newblock {\em Deep Sea Research\/}~{\em 3}, 152--153.

\bibitem[\protect\citeauthoryear{Timmermans, Toole, Krishfield, and
  Winsor}{Timmermans et~al.}{2008}]{Timmermanns2008}
Timmermans, M.-L., J.~Toole, R.~Krishfield, and P.~Winsor (2008).
\newblock Ice-tethered profiler observations of the double-diffusive staircase
  in the canada basin thermocline.
\newblock {\em Journal of Geophysical Research: Oceans\/}~{\em 113\/}(C1).

\bibitem[\protect\citeauthoryear{Turner}{Turner}{2010}]{turner2010}
Turner, J.~S. (2010).
\newblock {The Melting of Ice in the Arctic Ocean: The Influence of
  Double-Diffusive Transport of Heat from Below}.
\newblock {\em Journal of Physical Oceanography\/}~{\em 40\/}(1), 249--256.

\bibitem[\protect\citeauthoryear{Wood, Garaud, and Stellmach}{Wood
  et~al.}{2013}]{wood_2013}
Wood, T.~S., P.~Garaud, and S.~Stellmach (2013).
\newblock A new model for mixing by double-diffusive convection
  (semi-convection). ii. the transport of heat and composition through layers.
\newblock {\em The Astrophysical Journal\/}~{\em 768\/}(2), 157.

\bibitem[\protect\citeauthoryear{Zaussinger}{Zaussinger}{2010}]{zaussinger_dis%
s}
Zaussinger, F. (2010).
\newblock {\em Numerical simulation of double-diffusive convection}.
\newblock Ph.\ D. thesis, University of Vienna.

\bibitem[\protect\citeauthoryear{{Zaussinger}, {Kupka}, and
  {Muthsam}}{{Zaussinger} et~al.}{2013}]{zaussinger_kupka_muthsam_2012}
{Zaussinger}, F., F.~{Kupka}, and H.~J. {Muthsam} (2013).
\newblock {Semi-convection}.
\newblock In M.~{Goupil}, K.~{Belkacem}, C.~{Neiner}, F.~{Ligni{\`e}res}, and
  J.~J. {Green} (Eds.), {\em Lecture Notes in Physics}, Volume 865 of {\em
  Lecture Notes in Physics}, pp.\  219. Springer.

\bibitem[\protect\citeauthoryear{{Zaussinger} and {Spruit}}{{Zaussinger} and
  {Spruit}}{2013}]{zaussinger_scn_2013}
{Zaussinger}, F. and H.~C. {Spruit} (2013).
\newblock {Semiconvection: numerical simulations}.
\newblock {\em Astronomy \& Astrophysics\/}~{\em 554}, A119.

\end{thebibliography}

\end{document}